\documentclass[pdflatex,sn-mathphys-num]{sn-jnl}% Math and Physical Sciences Numbered Reference Style
\usepackage{graphicx}%
\usepackage{multirow}%
\usepackage{amsmath,amssymb,amsfonts}%
\usepackage{amsthm}%
\usepackage{mathrsfs}%
\usepackage[title]{appendix}%
\usepackage{xcolor}%
\usepackage[normalem]{ulem}%
\usepackage{cancel}%
\usepackage{textcomp}%
\usepackage{manyfoot}%
\usepackage{booktabs}%
\usepackage{algorithm}%
\usepackage{algorithmicx}%
\usepackage{algpseudocode}%
\usepackage{listings}%
\usepackage{comment}
\usepackage{subcaption}
\usepackage{gensymb}
\usepackage{tabularx}
\usepackage{adjustbox}

\usepackage{siunitx}
\usepackage{lineno}
\newcommand{\tr}{\operatorname{tr}}

\newcommand{\DEL}[1]{\ifmmode\textcolor{teal}{\cancel{#1}}\else\textcolor{teal}{\sout{#1}}\fi}

\theoremstyle{thmstyletwo}%

\theoremstyle{thmstylethree}%

\begin{document}

\title[Article Title]{Verification of a sequential thermo-poroelasticity formulation in PFLOTRAN}

%{Formulation and verification of sequentially coupled hydrothermal flow and geomechanical processes
%%=============================================================%%
%% GivenName	-> \fnm{Joergen W.}
%% Particle	-> \spfx{van der} -> surname prefix
%% FamilyName	-> \sur{Ploeg}
%% Suffix	-> \sfx{IV}
%% \author*[1,2]{\fnm{Joergen W.} \spfx{van der} \sur{Ploeg}
%%  \sfx{IV}}\email{iauthor@gmail.com}
%%=============================================================%%

\author*[1]{\fnm{J.} \sur{Al Kubaisy}} \email{jumanah.alkubaisy@pnnl.gov} % \orcidlink{0000-0002-7197-0691}
\author[1]{\fnm{G. E.} \sur{Hammond}}
\author[2]{\fnm{S.} \sur{Karra}}
\author[1]{\fnm{J.} \sur{Burghardt}}
\author[3]{\fnm{L.} \sur{Murdoch}}
\author[1]{\fnm{T.} \sur{Johnson}}
\author[2]{\fnm{K. M.} \sur{Rosso}}

\affil*[1]{\orgdiv{Energy and Environment Directorate}, \orgname{Pacific Northwest National Laboratory}, \orgaddress{\city{Richland}, \postcode{99352}, \state{WA}, \country{USA}}}

\affil[2]{\orgdiv{Integrated Discovery Sciences Directorate}, \orgname{Pacific Northwest National Laboratory}, \orgaddress{\city{Richland}, \postcode{99352}, \state{WA}, \country{USA}}}

\affil[3]{\orgdiv{Environmental Engineering and Earth Science Department}, \orgname{Clemson University}, \orgaddress{\city{Clemson}, \postcode{29634}, \state{SC}, \country{USA}}}

%\affil[4]{\orgdiv{Integrated Discovery Sciences Directorate}, \orgname{Pacific Northwest National Laboratory}, \orgaddress{\city{Richland}, \postcode{99352}, \state{WA}, \country{USA}}}

%%==================================%%
%% Sample for unstructured abstract %%
%%==================================%%

\abstract{
We present the verification of a thermo--hydrologic--mechanical capability implemented within the PFLOTRAN framework, with emphasis on benchmark-based assessment of the THM implementation. The thermal–hydrologic (TH) equations for mass and energy balance are solved on control-volume blocks or Voronoi cells, while the quasi-static momentum balance is solved on an element-based dual mesh. The coupling is achieved using a strictly sequential, non-iterative fixed-stress split strategy in which the TH system is solved implicitly for pressure and temperature, followed by a mechanics update for the displacement unknowns. Several verification problems are set up against poroelastic and thermo-poroelastic benchmarks, demonstrating agreement with analytical or semi-analytical benchmark responses for pressure diffusion, the temperature field, and mechanical deformation. In addition, we propose a treatment for discontinuities (e.g., fractures) based on mapping between mechanical and flow degrees of freedom, and validate the approach by comparison to an analytical solution. This work establishes the basis for thermo–poroelastic coupling in PFLOTRAN and provides a solid modeling foundation for a range of applications (e.g., enhanced geothermal systems and other subsurface energy storage) involving coupled thermal–hydrologic–mechanical (THM) processes in geologic porous media.
}


\maketitle

\section{Introduction}\label{sec:intro}
Coupled thermal–hydrologic–mechanical (THM) processes in geological porous media govern the evolution of pressure, temperature, stress, and deformation across a wide range of subsurface applications, including underground energy storage \citep{Heinemann_2021}, carbon sequestration \citep{Rutqvist_2012}, and geothermal energy systems \citep{McClure_2012}. In such systems, pore pressure changes can impact mechanical stresses resulting in compaction, stress redistribution, and fracture opening and closure \citep{biot_1941, Biot1955, Wang_2000}. Temperature variations introduce thermoelastic effects and can strongly affect fluid and rock properties, with corresponding impacts on flow and pressure diffusion \citep{coussy_2004, LewisSchrefler1998}. In many settings, geochemical coupling through thermo–hydrologic–mechanical–chemical (THMC) processes introduces additional feedback via fluid–rock interactions (e.g., dissolution–precipitation and mineral alteration). These reactions can alter porosity and permeability \citep{STEEFEL_2005, Rutqvist_2003} and may also affect mechanical strength and stiffness, thereby influencing flow, heat transport, and deformation. Capturing these dynamics is important for understanding and assessing performance in engineered subsurface operations and natural hydrothermal environments.

Robust numerical approximations of THM coupled processes remain challenging because they combine nonlinear flow and heat transport with mechanical deformation over highly heterogeneous media and multiple spatial and temporal scales. Monolithic formulations where a single system of equations is set up to solve all primary unknowns simultaneously can offer strong coupling but may be complex to implement and computationally demanding--particularly for large three-dimensional problems-- while sequential (or operator-splitting) strategies improve modularity but must be designed to preserve stability and accuracy \citep{LewisSchrefler1998, kim_2011}. In the latter approach, the coupled system is decomposed into subproblems (typically a flow solve and a quasi-static momentum-balance solve) that are advanced sequentially within each time step, exchanging pressure, temperature, and volumetric strain to enforce coupling.

Two common splits are the fixed-stress split, in which the flow problem is solved while holding the total mean stress fixed, and the undrained split, in which the mechanics problem is solved first assuming no change in fluid mass content and then the flow equation is solved sequentially to relax that constraint and update the pressure over the full time step. Among these, the fixed-stress split is frequently preferred due to its favorable stability and convergence behavior for poroelastic systems \citep{kim_2011, SettariMourits1998, MikelicWheeler2013}.

Several open-source codes support coupled THM processes for subsurface applications—such as GEOSX \citep{Settgast2024}, OpenGeoSys \citep{ogs:6.5.6}, and MOOSE \citep{harbour2025moose}. A key advantage of open-source software is transparency in its implementation, along with the ability to modify and extend governing equations, constitutive models, and coupling strategies to address application-specific requirements and modeling challenges. In this context, there is strong motivation to develop THM capabilities within established community codes that already provide robust, scalable solvers for flow and transport on high-performance computing platforms. Accordingly, our work focuses on PFLOTRAN, extending its thermal–hydrologic and reactive-transport infrastructure to enable robust coupled THM simulations.

PFLOTRAN is an open-source, massively parallel code for modeling subsurface flow, heat transport, and reactive transport on high-performance computing platforms \citep{hammond_2014, lichtner_2015, lichtner_2001}. Over the past two decades, it has established a robust foundation for simulating coupled thermal–hydrologic–chemical (THC) processes in geologic media. A geomechanics capability was introduced to PFLOTRAN over a decade ago; however, the coupling remained limited in scope and lacked the robustness required for fully coupled thermo–hydrologic–mechanical (THM) simulations.

The primary objective of this manuscript is to verify the PFLOTRAN THM implementation against benchmark problems that exercise pressure diffusion, thermo-poroelastic response, deformation, discontinuity loading, and borehole stress redistribution. More specifically, we aim to: (i) document the benchmark suite used to assess the THM capability; (ii) verify the implementation against analytical or semi-analytical solutions for pressure, temperature, deformation, stress, and crack-opening response; and (iii) demonstrate the consistency of the mapping between flow and geomechanics degrees of freedom for problems involving discontinuities such as fractures or pressurized cracks.

In this work, we present the formulation needed to define the verification problems and the corresponding benchmark results for a thermo–poroelasticity capability implemented within the PFLOTRAN framework. The thermal–hydrologic (TH) equations for mass and energy balance are solved using the finite-volume method, and the quasi-static momentum balance for the mechanics problem is solved using the finite-element method. The coupling is achieved via a strictly sequential, non-iterative fixed-stress split: the TH system is solved implicitly for pressure and temperature, followed by a mechanics update for displacement unknowns, consistent with standard poromechanics coupling theory \citep{biot_1941, coussy_2004, kim_2011}.

\section{Approach}\label{sec:method}
\subsection{Governing equations}\label{ssec:gov_eqn}
%th equations: mass and energy balance
We consider coupled flow and mechanics for single-phase, variably saturated, non-isothermal systems under a small-strain, quasi-static assumption. Unless stated otherwise for a specific benchmark, source terms are represented by $Q_w$ and $Q_e$, and gravity enters through the Darcy and body-force terms below.  In thermal-hydraulic (TH) flow mode, the mass balance is given by
\begin{equation}
\frac{\partial (\phi S \rho)}{\partial t} + \nabla \cdot (\rho \mathbf{q}) = Q_w,
 \label{eq:mass_balance}
\end{equation}
here $\phi$ is the porosity, $S$ is the saturation, $\rho$ is the fluid mass density, $t$ is time, $\mathbf{q}$ is the volumetric flux, and $Q_w$ is the mass source term. The volumetric flux is calculated using Darcy's law
\begin{equation}
\mathbf{q} = - \frac{k k_r}{\mu} \nabla(p -\rho g z),
\label{eq:darcy_equation}
\end{equation}
where $k$ is the intrinsic permeability, $k_r$ is the relative permeability, $\mu$ is the dynamic fluid viscosity, $p$ is the pressure unknown, $g$ is the gravity acceleration, and $z$ is the vertical coordinate. The energy balance of the system is described by
\begin{equation}
\frac{\partial}{\partial t}
\Big(
\phi S \rho U + (1-\phi)\rho_r c_p T
\Big) +
\nabla \cdot (\rho \mathbf{q} H - \kappa \nabla T) =
 Q_e,
 \label{eq:energy_balance}
\end{equation}
where $U$ is the fluid internal energy, $\rho_r$ is the rock (solid skeleton) density, $c_p$ is the effective rock heat capacity used in the PFLOTRAN TH formulation, $T$ is the temperature unknown, $H$ is the fluid enthalpy, $\kappa$ is the thermal conductivity combining both fluid and porous medium contributions, and $Q_e$ is the energy source term.

%Momentum eqn.
The mechanical deformation is governed by the quasi-static balance of linear momentum
\begin{equation}
\nabla \cdot \boldsymbol{\sigma} + \rho_b \boldsymbol{g} = \boldsymbol{0}.
\label{eq:balance_momentum}
\end{equation}
where $\boldsymbol{\sigma}$ is the Cauchy total stress tensor, $\boldsymbol{g}$ is the gravity vector (or, more generally, a prescribed specific body force), and $\rho_b$ is the bulk density calculated by
\begin{equation}
\rho_b = \phi \rho + (1-\phi) \rho_r.
\label{eq:bulk_density}
\end{equation}
The formulation in Eq.~\eqref{eq:balance_momentum} assumes negligible inertial effects (i.e., quasi-static conditions), infinitesimal strains, and the absence of external body forces other than gravity. The stress-strain relationship is
\begin{equation}
\boldsymbol{\sigma} = \lambda \tr(\boldsymbol{\varepsilon}) \boldsymbol{I} +
                      2 \mu \boldsymbol{\varepsilon} -
                    b p \boldsymbol{I} -
                    \beta_T \Delta T \, \boldsymbol{I},
\label{eq:stress_strain}
\end{equation}
$\lambda$ is the Lamé modulus and $\mu$ is the shear modulus. $\boldsymbol{I}$ is the identity matrix and $\tr(\boldsymbol{\varepsilon})$ is the trace of the strain tensor. The last two terms on the right-hand side include Biot's theory \citep{biot_1941} and thermal effects \citep{coussy_2004}; $b$ is Biot's coefficient, $\Delta T=T-T^0$ is the temperature change relative to the reference temperature $T^0$, and $\beta_T$ is the isotropic thermal-stress coefficient (for example, $\beta_T=3K_{dr}\alpha_T$ when a linear thermal-expansion coefficient $\alpha_T$ is introduced explicitly). Throughout this work, we adopt the convention that positive stresses denote tension and negative stresses denote compression. Biot's coefficient \citep{biot_1941} is calculated by
% Biot's coefficient
\begin{equation}
b = 1 - \frac{K_{dr}}{K_s},
\label{eq:biots_coeff}
\end{equation}
where $K_{dr}$ is the drained bulk modulus and $K_{s}$ the bulk modulus of the solid grain. The strain-displacement relationship under the small deformation assumption is given by
\begin{equation}
\mathbf{\varepsilon}=\tfrac12(\nabla \mathbf{u}+(\nabla \mathbf{u})^{T}),
\label{eq:strain_displacement}
\end{equation}
where $\nabla \mathbf{u}$ is the displacement gradient and $(\cdot)^T$ denotes transpose. The volumetric strain, which represents the relative change in bulk volume, is defined as the trace of the strain tensor:
\begin{equation}
\varepsilon_v = \mathrm{tr}(\boldsymbol{\varepsilon}).
\label{eq:strain_vol}
\end{equation}
\subsection{Finite volume method for thermal-hydraulic (TH) model}\label{ssec:discretization}
\begin{figure}
  \centering
  \begin{subfigure}[t]{0.5\textwidth}
    \centering
    \includegraphics[scale=0.6,trim={9cm 4cm 8cm 4cm},clip]{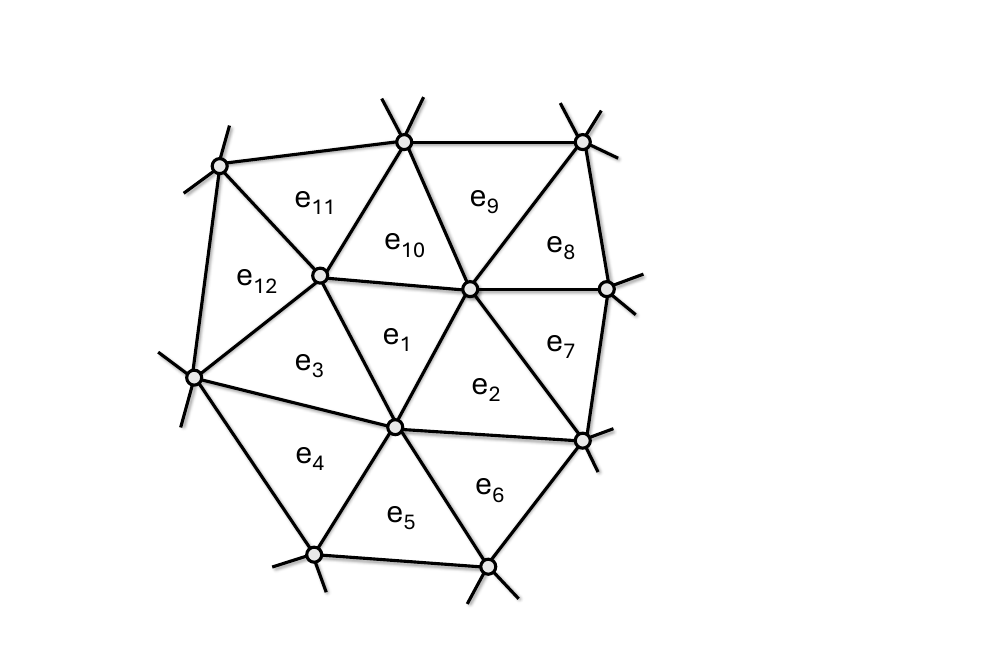}
    \caption{}
    \label{fig:ele_mesh}
  \end{subfigure}\hfill
  \begin{subfigure}[t]{0.5\textwidth}
    \centering
    \includegraphics[scale=0.6,trim={9cm 4cm 8cm 4cm},clip]{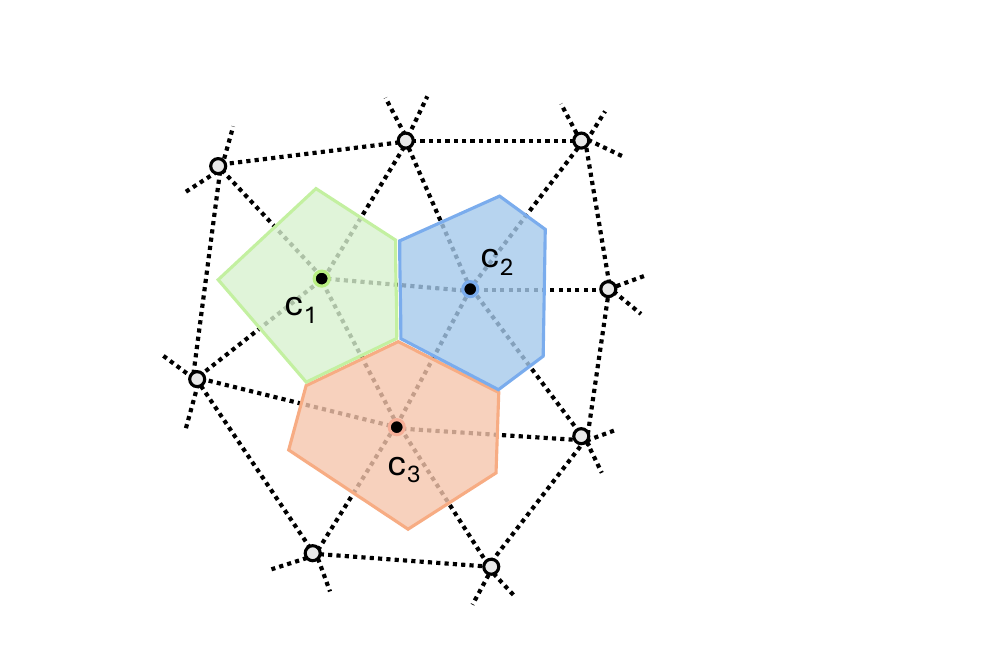}
    \caption{}
    \label{fig:voronoi_mesh}
  \end{subfigure}
  \caption{Two-dimensional illustration of dual-mesh approach: (a) element-based mesh with triangular elements ($e_1$-$e_{12}$), and (b) generated 2D Voronoi cells ($c_1$-$c_3$).}
  \label{fig:dual_mesh}
\end{figure}
\begin{figure}[t]
    \centering
    \includegraphics[scale=0.55,trim={5cm 4.5cm 5.5cm 5cm},clip]{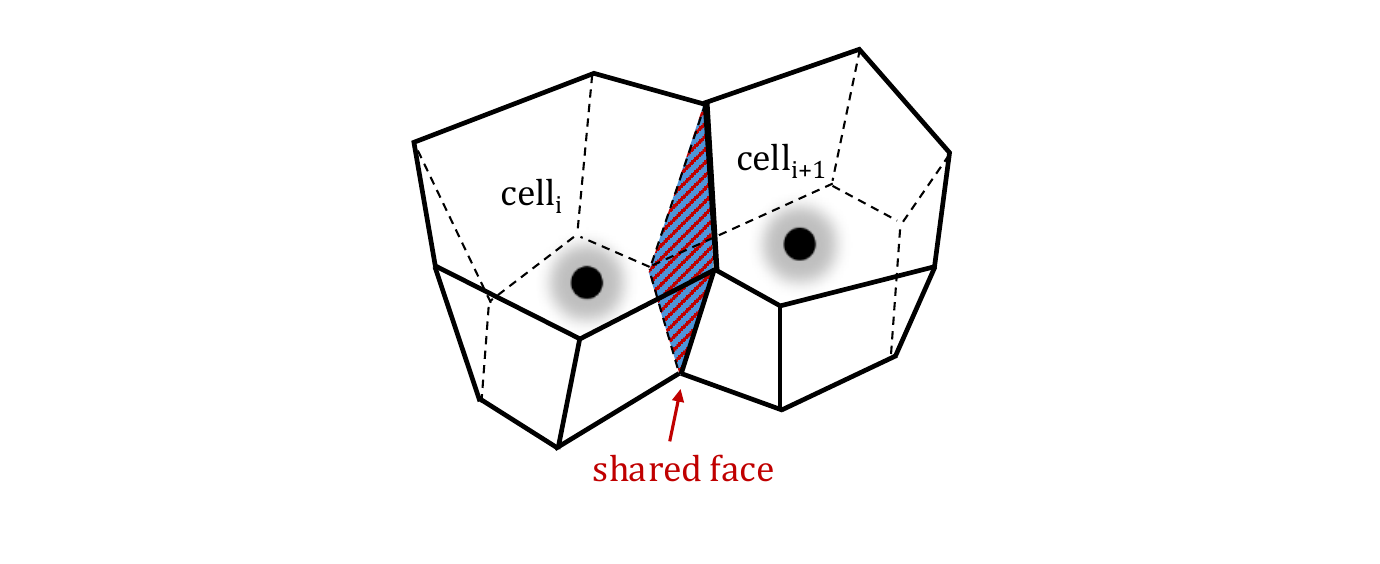}
    \caption{Three-dimensional schematic of two neighboring Voronoi control volumes, $\mathrm{cell}_i$ and $\mathrm{cell}_{i+1}$, sharing a common face across which intercell fluxes are evaluated.}
    \label{fig:2_voronoi_cells}
\end{figure}

The flow variables, pressure and temperature, are defined cell-wise when solving the mass balance Eq.~\eqref{eq:mass_balance} and the energy balance Eq.~\eqref{eq:energy_balance}. The control volume cells can be structured blocks or represented using Voronoi-type \citep{Okabe_2000, Voronoi_1908, Delaunay_1934}; see Fig. \ref{fig:voronoi_mesh}. Voronoi cells are widely used to construct unstructured meshes \citep{Lie_2019, Aavatsmark_2002, Palagi_1994} that are particularly well-suited for finite-volume discretizations of flow and transport. We briefly note that the construction of such cells follows the principle that the line connecting two neighboring cell centers is orthogonal to their shared face. This orthogonality condition is crucial for employing the two‑point flux approximation (TPFA) \citep{Edwards_1998,Eymard_2000, Aavatsmark_2002}. Fig. \ref{fig:2_voronoi_cells} shows two neighboring Voronoi cells ($\mathrm{cell}_i$ and $\mathrm{cell}_{i+1}$) highlighting the shared face.

The finite-volume formulation is obtained by (i) integrating the mass balance Eq.~\eqref{eq:mass_balance} over a control volume $V_i$, (ii) applying the divergence theorem to convert the flux divergence into a boundary flux integral, and (iii) expressing the boundary integral as a sum of fluxes over the faces $f\in \partial V_i$. This results in the following semi-discrete finite-volume approximation:
\begin{equation}
V_i\,\frac{\mathrm{d}}{\mathrm{d}t}\left(\phi_i S_i \rho_i\right)
\;+\;
\sum_{f\in\partial V_i} F_{i,f}
\;=\;
V_i\,Q_{w,i},
\label{eq:fvm_semidiscrete}
\end{equation}
where $F_{i,f}$ denotes the face mass flux. The same approach is used to derive the semi-discrete finite-volume approximation of the energy-balance equation, Eq.~\eqref{eq:energy_balance}:
%the finite-volume formulation for the energy balance equation, Eqn. \eqref{eq:energy_balance}

\begin{equation}
V_i\,\frac{\mathrm{d}}{\mathrm{d}t}
\left[
\Big(\phi_i S_i \rho_i U_i\Big)
+
\Big((1-\phi_i)\rho_r c_p T_i\Big)
\right]
\;+\;
\sum_{f\in\partial V_i} F^{E}_{i,f}
\;=\;
V_i\,Q_{e,i},
\label{eq:fv_energy_semidiscrete}
\end{equation}
with $F^{E}_{i,f}$ denoting the total energy flux across face $f$.
\subsection{Finite element method for mechanics model}
The mechanics model uses an element-based mesh, which also serves as the basis for constructing the Voronoi mesh, see Fig. \ref{fig:dual_mesh}. PFLOTRAN currently supports two element types for coupled THM simulations: hexahedral and tetrahedral elements.

Let $\Omega\subset\mathbb{R}^d$ with boundary $\partial\Omega=\Gamma_u\cup\Gamma_t$, where displacement ($\mathbf{u}=\bar{\mathbf{u}}$) is prescribed on $\Gamma_u$ and traction ($\boldsymbol{\sigma}\mathbf{n}=\bar{\mathbf{t}}$) is prescribed on $\Gamma_t$. We derive the weak form by multiplying Eq.~\eqref{eq:balance_momentum} with a test function $\mathbf{w}$ that vanishes on the boundaries, integrate over $\Omega$, and apply integration by parts: find $\mathbf{u}\in\mathcal{U}$ such that for all $\mathbf{w}\in\mathcal{V}$,
\begin{equation}
\int_{\Omega}
\boldsymbol{\sigma}(\mathbf{u},p,T):\nabla^{s}\mathbf{w}\,\mathrm{d}\Omega
=
\int_{\Omega}\rho_b\mathbf{g}\cdot\mathbf{w}\,\mathrm{d}\Omega
+
\int_{\Gamma_t}\bar{\mathbf{t}}\cdot\mathbf{w}\,\mathrm{d}\Gamma,
\label{eq:weak_momentum_u}
\end{equation}
where the trial space ($\mathcal{U}$) and the test space ($\mathcal{V}$) are defined as
\begin{equation}
\mathcal{U}=\{\mathbf{u}\in[H^1(\Omega)]^d:\mathbf{u}=\bar{\mathbf{u}}\ \text{on}\ \Gamma_u\},
\qquad
\mathcal{V}=\{\mathbf{w}\in[H^1(\Omega)]^d:\mathbf{w}=\mathbf{0}\ \text{on}\ \Gamma_u\}.
\end{equation}

For a linear finite element approximation, the displacement field is interpolated from nodal values using linear (Lagrange) shape functions. On an element $e$ with $n$ nodes,
\begin{equation}
\mathbf{u}^h(\mathbf{x}) = \sum_{a=1}^{n} N_a(\mathbf{x})\,\mathbf{u}_a,
\label{eq:fe_u_interp}
\end{equation}
where $N_a(\mathbf{x})$ is a linear shape function and $\mathbf{u}_a$ is the nodal displacement vector.
%+ mention the stress and strain approximation and averaging\\
%\paragraph{Representing material discontinuities}

\section{Sequentially coupled solution strategy}\label{ssec:pmc}
\subsection{Flow field initialization}
In PFLOTRAN's thermal–hydrologic (TH) flow mode, the flow model is initialized by prescribing initial conditions for the primary variables—pressure ($p^0$) and temperature ($T^0$)—over the domain. In practice, these fields are often initialized from site-specific datasets derived from field observations (e.g., pressure measurements and temperature logs) and may be represented as spatially varying distributions, such as a hydrostatic pressure profile defined by a reference pressure and gravitational potential and a temperature field based on a geothermal gradient or other measured spatial trends.
\subsection{Mechanics initialization}
Once the flow field is set, we ensure that the mechanical equilibrium is met by solving an initial mechanics step \citep{kim_2011}. The volumetric strain for the reference state ($\varepsilon_v^0$) is stored for later computation.
\subsection{Fixed-stress split}
\label{ssec:fstress_split}
The solution approach here is based on solving the flow problem first, assuming the increment in total mean stress is fixed, i.e., $\sigma^{n+1}-\sigma^{n} = \sigma^{n}-\sigma^{n-1}$ \citep{KimTchelepiJuanes2011}. Within each flow iteration ($k$) at the current timestep ($n+1$), we solve for the flow unknowns, pressure and temperature. The porosity ($\phi$) is the only variable dependent on both flow (via $p$ and $T$) and mechanics (via $\varepsilon_v$, $K_{dr}$, and $K_s$) where it is updated using
\begin{equation}
\begin{split}
\phi^{n+1, k}(p^n, p^{n+1,k}, T^{n+1,k}) = &\phi^0 + b (\varepsilon_v^n -\varepsilon_v^0) +
\frac{b - \phi^0}{K_s} (p^{n}-p^0) + \\
&\Bigg(\frac{b^2}{K_{dr}} + \frac{b - \phi^0}{K_s} \Bigg) (p^{n+1,k}-p^n) +
\alpha_{\phi} (T^{n+1,k} - T^0),
 \label{eq:porosity_flow}
 \end{split}
\end{equation}
here the superscripts $0$ and $n$ denote the initial value and the value at the previous timestep, respectively, whereas the superscript $(n+1,k)$ denotes the value at the current timestep evaluated at the $k$th iteration. This method adopts the porosity update given by \cite{Burghardt_2017} for a strictly sequential coupling approach and incorporates the thermal effects into the formulation. The reference temperature ($T^0$) and pressure ($p^0$) are taken from the initialized TH state, and $\alpha_{\phi}$ has units of inverse temperature, so that the product $\alpha_{\phi}(T^{n+1,k}-T^0)$ is dimensionless and can be added consistently to the porosity. Under this sign convention, a positive $\alpha_{\phi}$ implies that heating at fixed pressure and strain increases porosity, whereas cooling decreases it. In the current implementation, the same value is assigned to $\alpha_{\phi}$ (Eq.~\ref{eq:porosity_flow}) and to $\alpha_{T}$, which enters through $\beta_T$ in Eq.~\ref{eq:stress_strain}. This procedure is iterated until the flow solution converges, after which the mechanics step follows, updating the displacement using the newly obtained pressure and temperature.

\subsection{Solution method for sequentially coupled THM}
The coupled thermo-poroelastic problem is solved using a sequential, non-iterative coupling scheme. The flow subproblem is treated fully implicitly and solved with a nonlinear Newton method, while the mechanical equilibrium (momentum balance) subproblem is solved as a linear system for incremental displacements. The coupling is enforced in the mass balance Eq.~\eqref{eq:mass_balance} and energy balance Eq.~\eqref{eq:energy_balance} by the porosity dependence on volumetric strain, mechanical properties, and pressure and temperature evolution in the system, see Eq.~\eqref{eq:porosity_flow}.

Over a time step $t^n \to t^{n+1}$, the nonlinear balance laws for mass and energy are solved in residual form,
\begin{equation}
\mathbf{R}\!\left(\mathbf{x}_{\mathrm{TH}}^{n+1};\,\varepsilon_v^n\right)=\mathbf{0},
\qquad
\mathbf{x}_{\mathrm{TH}}=\begin{bmatrix} p \\ T \end{bmatrix},
\label{eq:Rth}
\end{equation}
where $p$ and $T$ are the primary TH unknowns and $\varepsilon_v^n$ denotes volumetric strain, unchanged during the TH solve within a sequential coupling step. For completeness, we recall that Newton's method computes updates $\delta\mathbf{x}_{\mathrm{TH}}^{k}$ by solving the linearized system
\begin{equation}
\mathbf{J}^{k}\,\delta\mathbf{x}_{\mathrm{TH}}^{k} \;=\; -\mathbf{R}^{k},
\qquad
\mathbf{x}_{\mathrm{TH}}^{k+1}=\mathbf{x}_{\mathrm{TH}}^{k}+\delta\mathbf{x}_{\mathrm{TH}}^{k},
\label{eq:newton}
\end{equation}
with $\mathbf{J}^{k}=\partial \mathbf{R}/\partial \mathbf{x}_{\mathrm{TH}}$ assembled from the fully implicit finite-volume discretization. After convergence of the TH Newton iterations, the mechanics subproblem is updated by solving the quasi-static momentum balance using the converged pressure and temperature as loads through thermo-poroelastic coupling (Eq.~\eqref{eq:stress_strain}) which yields a linear system of the form
\begin{equation}
\mathbf{K}\,\mathbf{u}^{n+1} \;=\; \mathbf{f}_{\mathrm{ext}}^{n+1} \;+\; \mathbf{f}_{p}(p^{n+1}) \;+\; \mathbf{f}_{T}(T^{n+1}),
\label{eq:mechLin}
\end{equation}
where $\mathbf{K}$ is the elastic stiffness matrix, $\mathbf{f}_{\text{ext}}^{n+1}$ captures body forces and traction boundary conditions, and $\mathbf{f}_{p}$ and $\mathbf{f}_{T}$ represent the nodal forces induced by pore pressure (Biot term) and thermal strain, respectively. There is no dependence of $\mathbf{K}$ on $\mathbf{u}$ for linear elasticity, hence, the mechanics solve is performed as a linear system each time step using an iterative Krylov solver. This strictly sequential solution strategy for THM processes captures the nonlinear TH physics while providing an efficient linear mechanics update at each time step.

We briefly summarize how PFLOTRAN leverages PETSc (Portable, Extensible Toolkit for Scientific Computation) \citep{petsc-web-page, petsc-user-ref, petsc-efficient} in its solution workflow. PETSc provides scalable data structures and solvers for large sparse systems on distributed-memory architectures. PFLOTRAN employs PETSc's nonlinear solver infrastructure (SNES) for the Newton iterations and preconditioned Krylov subspace methods (KSP) for solving each linearized system, Eq.~\eqref{eq:newton}, as well as the mechanics system, Eq.~\eqref{eq:mechLin}. PETSc also manages parallel vector and matrix assembly, enabling the coupled THM simulations to scale efficiently on high-performance computing platforms \citep{hammond_2014}.

\section{Numerical experiments}\label{sec:num_exp}
\begin{figure}
    \centering
    \includegraphics[scale=0.58,trim={6.3cm 4.5cm 5cm 3cm},clip]{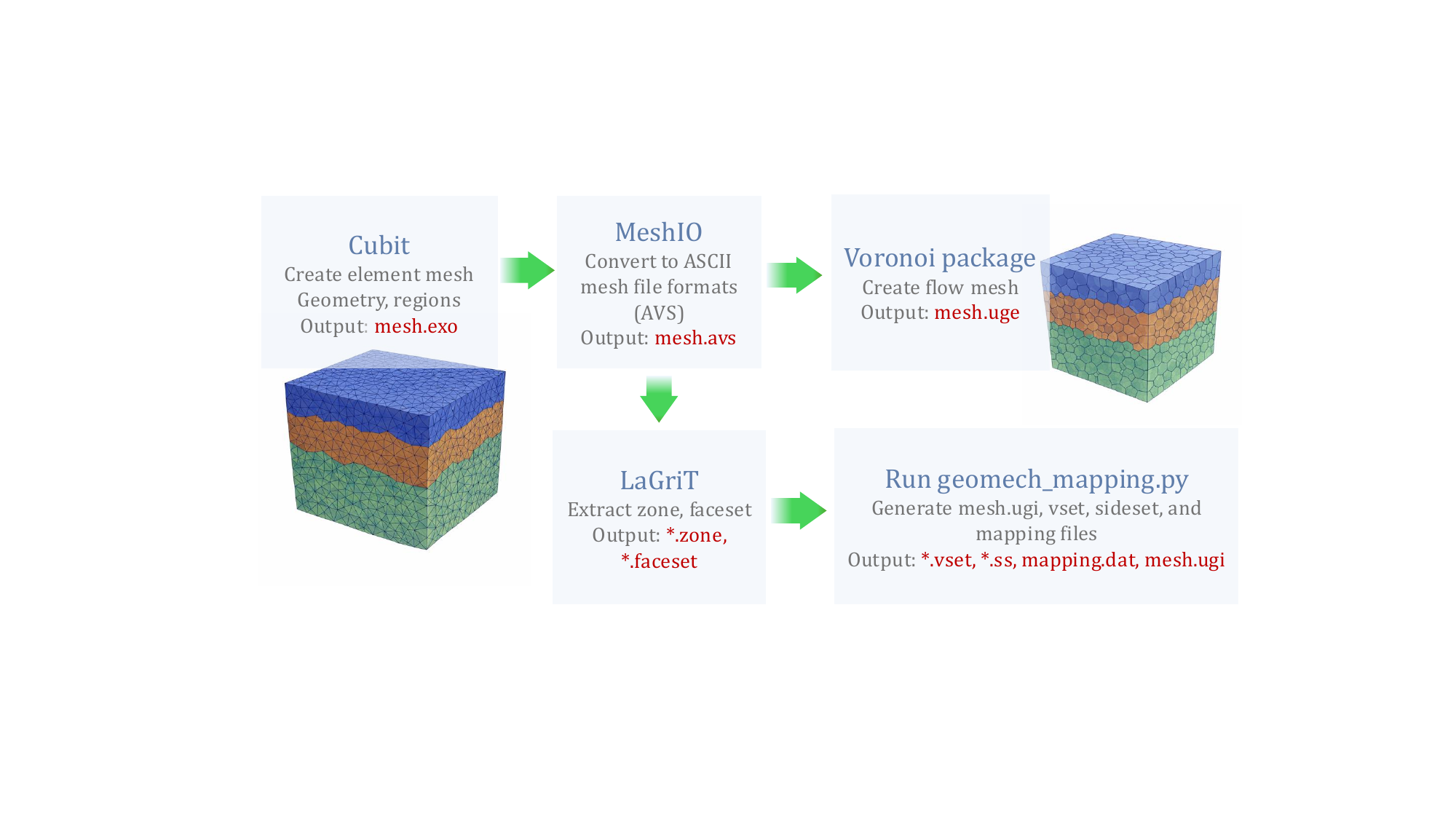}
    \caption{The meshing workflow used to generate unstructured grids and input files for coupled flow–mechanics simulations in PFLOTRAN. The workflow produces: (i) mesh.uge, the unstructured grid for flow simulations (cell list and connectivity); (ii) mesh.ugi, the standard finite-element mesh description (element vertex lists and vertex coordinates); (iii) *.vset, the vertex-set file used to define domain regions; (iv) *.ss, the sideset file specifying boundary nodes where stress conditions are applied; and (v) mapping.dat, the mapping between flow and geomechanics nodes (typically one-to-one).}
    \label{fig:mesh_workflow}
\end{figure}

We present a suite of benchmark problems to verify the THM implementation in PFLOTRAN. The verification experiments include Terzaghi’s consolidation problem \citep{Terzaghi_1943}, Schiffman’s problem \citep{Schiffman_1958}, Mandel’s problem \citep{Mandel_1953}, and the thermo-poroelastic bar problem of \citet{Bai_2005}. For each case, numerical results are compared with analytical or semi-analytical solutions for the transient evolution of pore pressure, temperature, deformation, or stress under prescribed loading and drainage conditions (applied to the mass balance Eq.~\eqref{eq:mass_balance}, energy balance Eq.~\eqref{eq:energy_balance} when relevant, and the momentum balance Eq.~\eqref{eq:balance_momentum}). We then demonstrate our approach for representing discontinuities (e.g., fractures or cracks). Finally, we verify the accuracy of the geomechanics implementation for circular borehole models.
To create unstructured meshes for the numerical experiment presented here, a combination of CUBIT \citep{cubit, cubit_16_2022}, LaGriT \citep{lagrit_software}, Voronoi \citep{voronoi_git}, and MeshIO python package \citep{meshio_github} are used in this work, see Fig. \ref{fig:mesh_workflow}. The structured grids (control volume blocks for flow and hexahedral elements for mechanics) are internally supported and generated by PFLOTRAN.

\subsection{1D consolidation problems}\label{ssec:1d_consolidation}
\begin{figure}
  \centering
  \begin{subfigure}[t]{0.4\textwidth}
    \centering
    \includegraphics[scale=0.55,trim={0.2cm 0cm 1cm 0cm},clip]{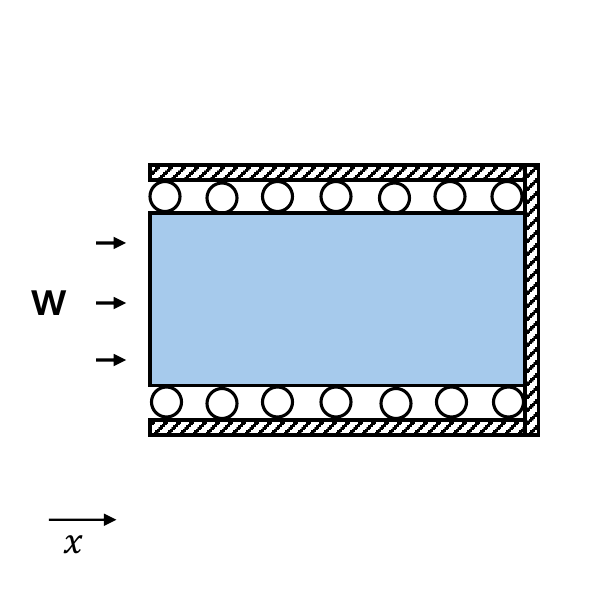}
    \caption{}
    \label{fig:1d_sketch}
  \end{subfigure}\hfill
  \begin{subfigure}[t]{0.55\textwidth}
    \centering
    \includegraphics[scale=0.15,trim={0cm 5cm 6cm 1cm},clip]{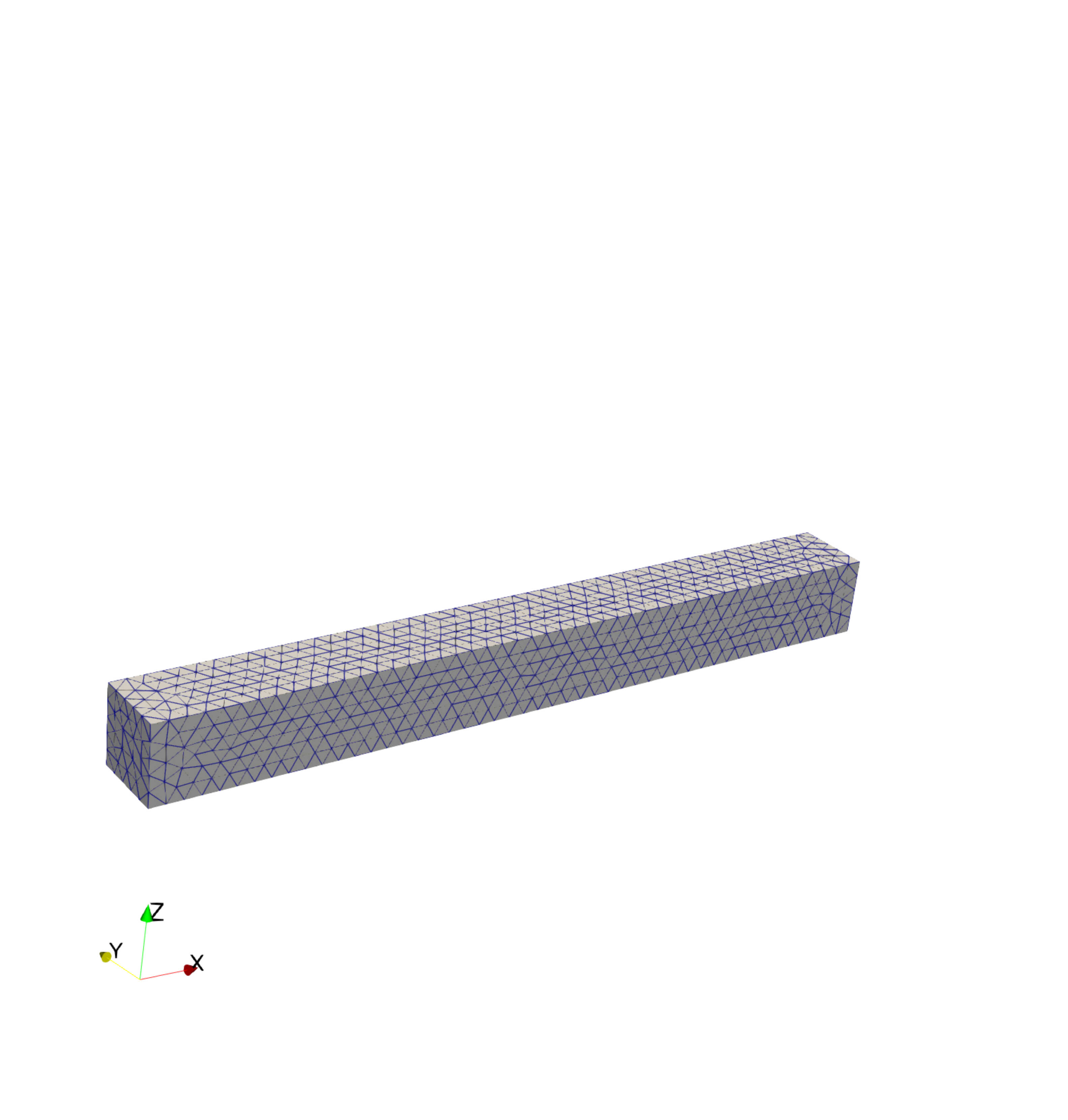}
    \caption{}
    \label{fig:1d_mesh}
  \end{subfigure}
  \caption{(a) Conceptual sketch of Terzaghi's problem and Schiffman's problem. A compressive stress load (\textbf{W}) is applied on the left face of a saturated porous medium, which also serves as the only drained boundary. The right boundary is fixed, while the top and bottom boundaries are prescribed with zero normal displacement (rollers). (b) Computational domain using tetrahedral elements.}
  \label{fig:1d_setup}
\end{figure}

We verify the coupled flow--geomechanics implementation using two classical one-dimensional consolidation benchmarks: Terzaghi's instantaneous (step) loading problem \citep{Terzaghi_1943} and the time dependent extension proposed by \citet{Schiffman_1958}. Both tests are set up as pseudo one-dimensional problems with domain length ($L_x$), with deformation and flow restricted to the horizontal direction $x\in[0,L_x]$. The primary unknowns are the pore pressure $p(x,t)$ and the displacement $u(x,t)$. Material properties are homogeneous and isotropic, and gravity is neglected for the verification cases (see Table \ref{tab:terzaghi_params} for model parameters, and Fig. \ref{fig:1d_sketch} and \ref{fig:1d_mesh} for conceptual model sketch and meshed domain, respectively); the boundary conditions are imposed on the mass balance Eq.~\eqref{eq:mass_balance} and the momentum balance Eq.~\eqref{eq:balance_momentum}.
%\nolinenumbers
\begin{table}[t]
  \centering
  \caption{Model setup and parameters in numerical experiment \ref{ssec:1d_consolidation} for both Terzaghi's problem and Schiffman's problem.}
  \label{tab:terzaghi_params}
  \begin{tabularx}{\linewidth}{X r l}
    \toprule
    Parameter & Value & Unit \\
    \midrule
    Domain length ($L_x$) & 1 & m \\
    Cross-sectional area ($L_y \times L_z$) & 0.01 & m$^{2}$ \\
    Young's modulus ($E$) & 100 & MPa \\
    Poisson's ratio ($\nu$) & 0.2 & -- \\
    Biot's coefficient ($b$) & 1 & -- \\
    Fluid bulk modulus ($K_f$) & $\infty$ & Pa \\
    Porosity ($\phi$) & 0.2 & -- \\
    Permeability ($k$) & \num{1e-8} & m$^{2}$ \\
    Fluid viscosity ($\mu$) & 1 & Pa$\cdot$s \\
    Applied compressive stress load, Terzaghi's setup ($\mathbf{w}$) & 1 & Pa \\
    Applied compressive stress load, Schiffman's setup ($\mathbf{w}$) & 3 & Pa \\
    Number of elements & 10.2k & -- \\
    Number of Voronoi cells & 2.4k & -- \\
    \bottomrule
  \end{tabularx}
\end{table}
%\linenumbers
\subsubsection*{Terzaghi's problem}\label{ssec:Terzaghi}
\begin{figure}
\centering
  \begin{subfigure}{\textwidth}
    \centering
    \includegraphics[scale=0.65,trim={0cm 0cm 0cm 0cm},clip]{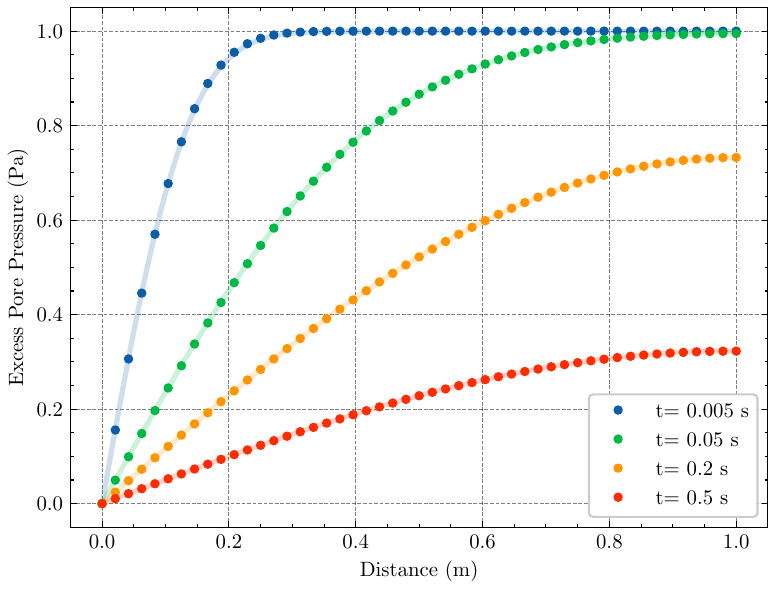}
    \caption{}
    \label{fig:terzaghi_press}
  \end{subfigure}\hfill
  \begin{subfigure}{\textwidth}
    \centering
    \includegraphics[scale=0.65,trim={0cm 0cm 0cm 0cm},clip]{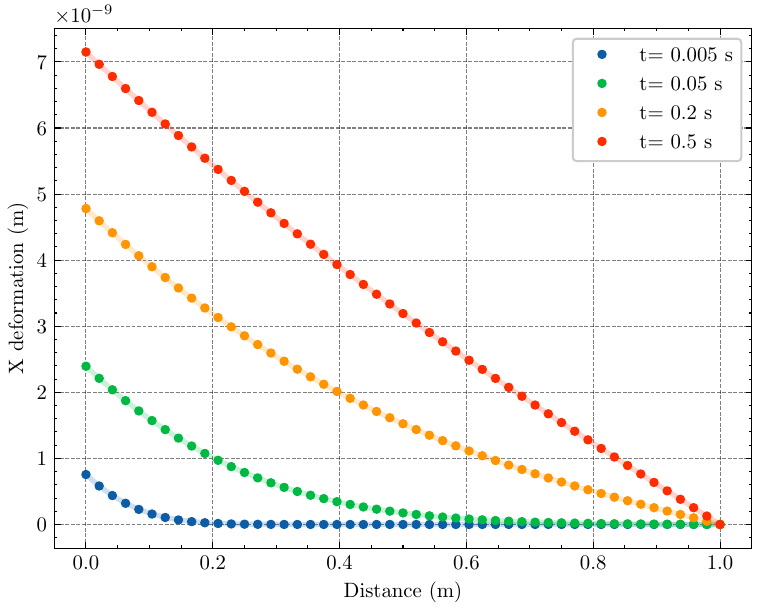}
    \caption{}
    \label{fig:terzaghi_disp}
  \end{subfigure}
\caption{Pore-pressure and displacement profiles for Terzaghi's problem in experiment~\ref{ssec:Terzaghi}. Symbols denote the PFLOTRAN solution and the solid line denotes the analytical solution.}
\label{fig:terzaghi_comparison}
\end{figure}

Terzaghi's classical consolidation problem \citep{Terzaghi_1943} considers an instantaneous application of a surface load \textbf{W} on a drained boundary (Dirichlet pressure condition for the mass balance Eq.~ \eqref{eq:mass_balance}), Fig. \ref{fig:1d_sketch}. Immediately after loading ($t\rightarrow 0^+$), the response is undrained and the load is balanced by a sudden increase in pore pressure. The pore pressure then dissipates due to drainage at $x=0$, effective stress increases, and the bar undergoes consolidation. The numerical solutions for pressure and displacement exhibit relative $L_2$-norm errors below 0.002 and 0.01, respectively (Table~\ref{tab:l2_err}), compared with the analytical solutions (Fig.~\ref{fig:terzaghi_comparison}).
\subsubsection*{Schiffman's problem}\label{ssec:Schiffman}
The problem proposed by \citet{Schiffman_1958} extends Terzaghi's consolidation model to time dependent loading, verifying the transient coupling between the flow and mechanical models. The domain geometry, initial conditions (for the mass balance Eq.~\eqref{eq:mass_balance} and energy balance Eq.~\eqref{eq:energy_balance} when applicable), and drainage conditions are the same as Terzaghi's problem; only the applied surface load differs. The load is increased linearly in time until reaching the maximum of $3 ~\mathrm{Pa}$ at $3~\mathrm{ms}$. During the ramp, pore pressure increases and diffuses simultaneously. This response differs from the purely undrained initial condition in Terzaghi's (step load) problem. After $3~\mathrm{ms}$, the load is held constant and the excess pore pressure dissipates as the bar undergoes consolidation.

\begin{figure}
  \begin{subfigure}{\textwidth}
    \centering
    \includegraphics[scale=0.65,trim={0cm 0cm 0cm 0cm},clip]{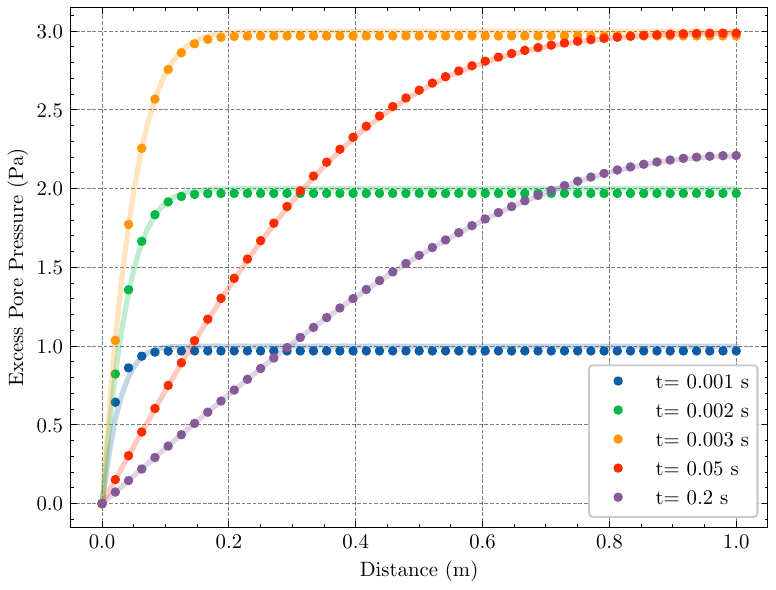}
    \caption{}
    \label{fig:schiffman_press}
  \end{subfigure}\hfill
  \begin{subfigure}{\textwidth}
    \centering
    \includegraphics[scale=0.65,trim={0cm 0cm 0cm 0cm},clip]{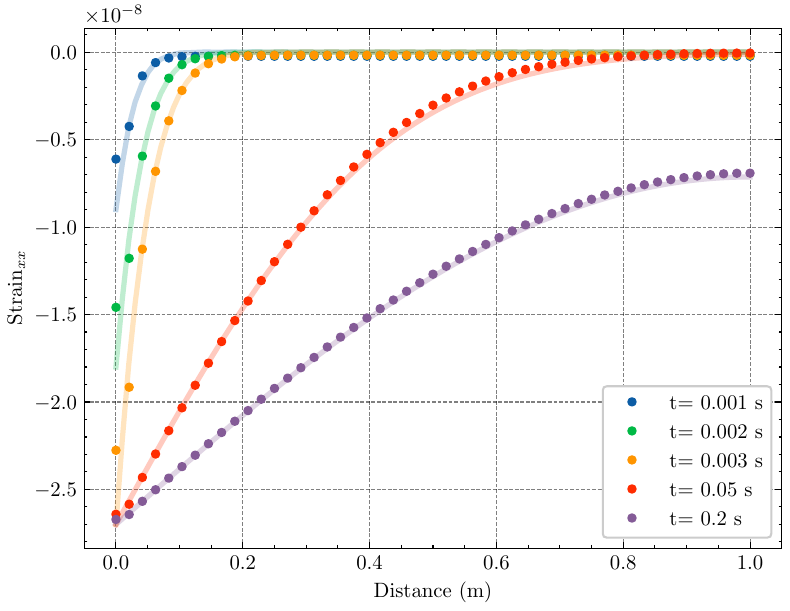}
    \caption{}
    \label{fig:schiffman_strain}
  \end{subfigure}
\caption{Pore-pressure and strain profiles for Schiffman's problem in experiment~\ref{ssec:Schiffman}. Symbols denote the PFLOTRAN solution and the solid line denotes the analytical solution.}
\label{fig:schiffman_comparison}
\end{figure}

Fig. \ref{fig:schiffman_comparison} presents the numerical approximations of pressure and strain alongside the analytical solutions \citep{Huang_2025}. The numerical results accurately capture (i) the pressure buildup during the loading phase, (ii) the subsequent pressure dissipation, and (iii) the evolution of principal strain in the system. The relative $L_2$-norm errors for pressure and strain are less than 0.04 and 0.32, respectively (Table \ref{tab:l2_err}).

\subsection{Mandel's problem} \label{ssec:mandel}
\begin{figure}[htbp]
  \centering
  \begin{subfigure}[t]{0.48\textwidth}
    \centering
    \includegraphics[scale=0.6]{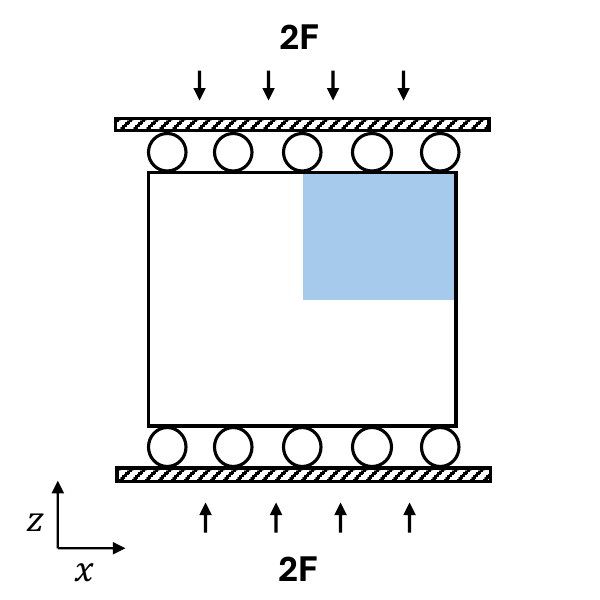}
    \caption{}
    \label{fig:conc_mandel}
  \end{subfigure}\hfill
  \begin{subfigure}[t]{0.48\textwidth}
    \centering
    \includegraphics[scale=0.15,trim={6cm 1cm 4cm 8cm},clip]{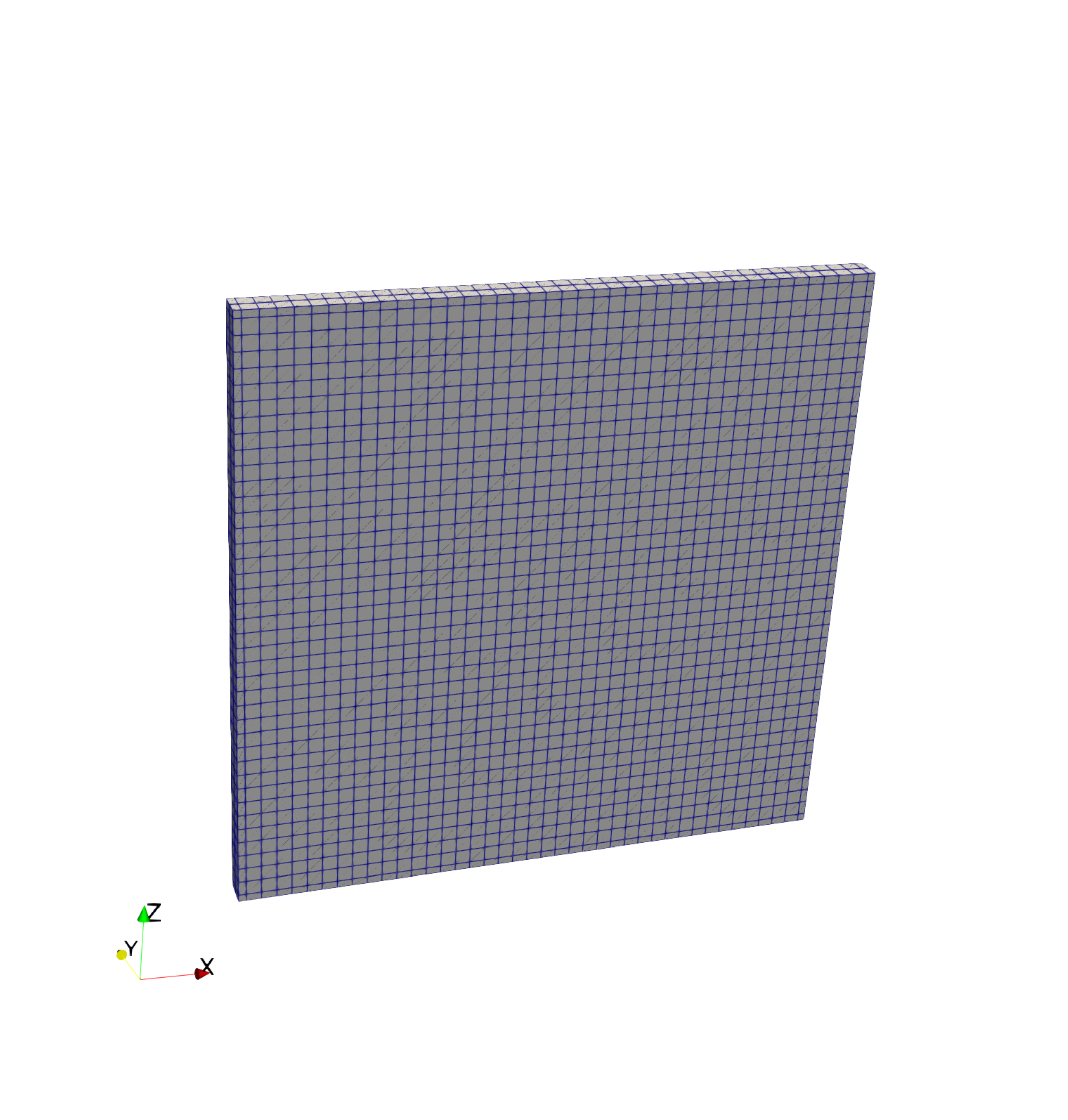}
    \caption{}
    \label{fig:domain_mandel}
  \end{subfigure}
  \caption{Setup of numerical experiment \ref{ssec:mandel}.
  (a) Conceptual sketch of Mandel's problem: a saturated domain is confined between two impermeable plates which are loaded by a constant compressive stress $(\textbf{2F})$ applied along the z-direction. The lateral boundaries ($x=\pm L_x$) are traction-free and fully drained. By symmetry of the geometry and loading, only the blue-shaded quarter-domain is modeled. Symmetry conditions are imposed along the $x=0$ and $z=0$ planes by prescribing zero normal displacement and zero normal fluid flux (roller and no-flow conditions). (b) Corresponding element mesh (hexahedral elements).}
  \label{fig:mandel_setup}
\end{figure}

Mandel’s problem \citep{Mandel_1953} is a classic poroelastic two-dimensional benchmark in which a rectangular, saturated porous medium is subjected to a suddenly applied compressive load \textbf{2F} (Fig. \ref{fig:conc_mandel}). In this experiment, we model one-quarter of the domain by exploiting geometric and loading symmetry about the mid-planes.

\begin{figure}[htbp]
\centering
  \begin{subfigure}[t]{\textwidth}
    \centering
    \includegraphics[scale=0.65,trim={0cm 0cm 0cm 0cm},clip]{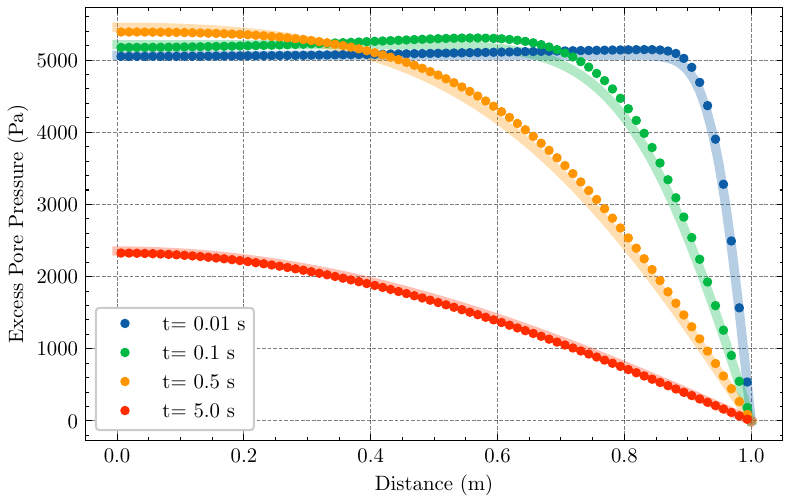}
    \caption{}
    \label{fig:mandel_pres}
  \end{subfigure}
  \begin{subfigure}[t]{\textwidth}
    \centering
    \includegraphics[scale=0.65,trim={0cm 0cm 0cm 0cm},clip]{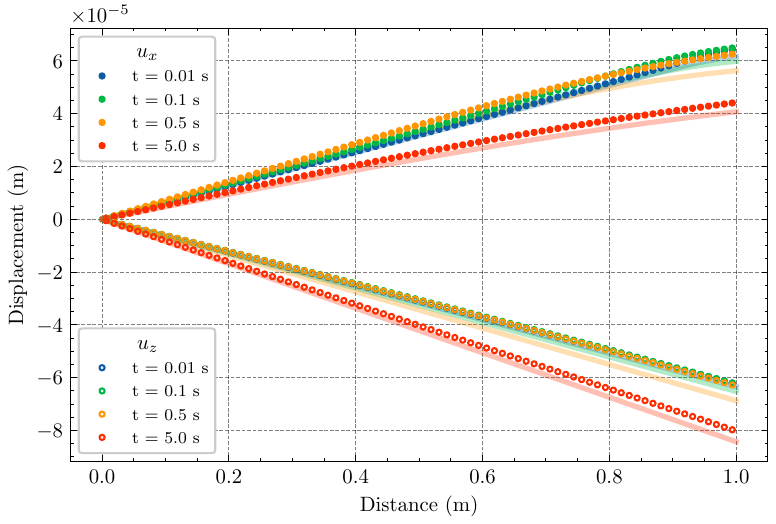}
    \caption{}
    \label{fig:mandel_disp}
  \end{subfigure}
\caption{Benchmark comparison for Mandel's problem in experiment~\ref{ssec:mandel}: (a) pore pressure profiles and (b) displacement profiles. Symbols denote the PFLOTRAN solution and the solid line denotes the semi-analytical solution. Positive displacement corresponds to the lateral direction (drained boundary at $x=1$ for the mass balance Eq.~\eqref{eq:mass_balance}), whereas negative displacement corresponds to the vertical direction (undrained boundary at $z=1$ for the momentum balance Eq.~\eqref{eq:balance_momentum}).}
\label{fig:mandel_press_disp}
\end{figure}

The analytical solutions for pressure, horizontal displacement, and vertical displacement, obtained from \cite{Castelletto_2015}, are plotted with the numerical results in Fig.~\ref{fig:mandel_press_disp}. Fig.~\ref{fig:mandel_pres} shows the pore pressure profile at several time instances, capturing the Mandel–Cryer effect — a non-monotonic pressure response arising from the coupling between fluid flow and mechanical deformation under mixed boundary conditions \citep{Abousleiman_1996}; drained boundaries corresponding to conditions on the mass balance Eq.~\eqref{eq:mass_balance}, traction-free and roller conditions on the momentum balance Eq.~\eqref{eq:balance_momentum}. Near the drained boundaries ($x=1$), fluid drains early and the local effective stress increases, causing the region to carry a progressively larger share of the applied load. To maintain global force equilibrium, total stress is redistributed toward the interior of the domain ($x=0$), where the material remains in a nearly undrained state. This transferred load is initially borne entirely by the pore fluid in the undrained interior, causing a temporary increase in pore pressure near $x=0$ that exceeds the initial undrained value. As the drainage front eventually reaches the center of the domain, the excess pore pressure dissipates and the system approaches its fully drained equilibrium.

Fig.~\ref{fig:mandel_disp} shows the horizontal displacement ($u_x$) and vertical displacement ($u_z$) profiles along the domain at the same time instances. The horizontal displacement reflects the lateral expansion of the domain as pore pressure dissipates and effective stresses redistribute. The vertical displacement $u_z$ is negative, indicating compaction under the applied load, with its magnitude increasing over time as the excess pore pressure dissipates. The relative $L_2$-errors for pressure, horizontal displacement, and vertical displacement are less than 0.03, 0.1, and 0.1, respectively (Table \ref{tab:l2_err}).

\begin{table}[htbp]
  \centering
  \caption{Model parameters for Mandel's problem \ref{ssec:mandel}.}
  \label{tab:mandel_params}
  \begin{tabularx}{\linewidth}{X r l}
    \toprule
    Parameter & Value & Unit \\
    \midrule
    Domain length, width, height ($L_x$, $L_y$, $L_z$) & 1.0, 0.025, 1.0 & m \\
    Young's modulus ($E$) & 100 & MPa \\
    Poisson's ratio ($\nu$) & 0.25 & -- \\
    Biot's coefficient ($b$) & 1 & -- \\
    Fluid bulk modulus ($K_f$) & $\infty$ & Pa \\
    Porosity ($\phi$) & 0.375 & -- \\
    Permeability ($k$) & \num{1e-12} & m$^{2}$ \\
    Fluid viscosity ($\mu$) & 0.001 & Pa$\cdot$s \\
    Applied compressive stress ($2\mathbf{F}$) & \num{1e4} & Pa \\
    Number of elements & 13k & -- \\
    Number of flow cells & 6.4k & -- \\
    \bottomrule
    \end{tabularx}
\end{table}

\subsection{\texorpdfstring{Thermo-poroelastic one-dimensional consolidation}{Thermo-poroelastic one-dimensional consolidation}} \label{ssec:thermo_bar}
\begin{figure}
  \centering
  \begin{subfigure}[t]{0.5\textwidth}
    \centering
    \includegraphics[scale=0.6]{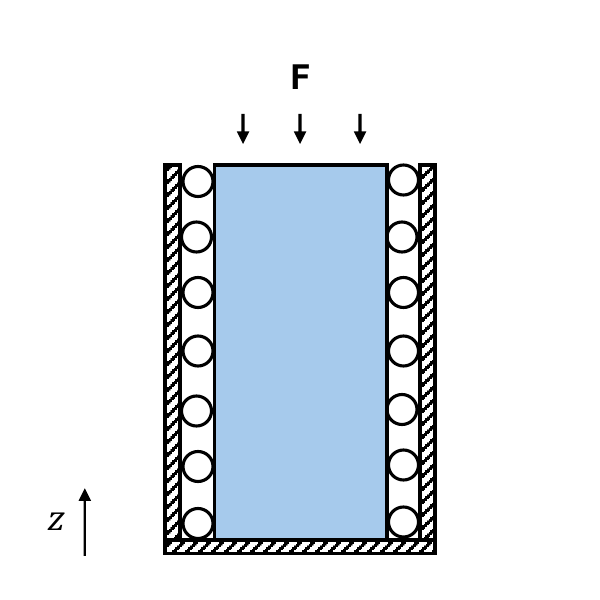}
    \caption{}
    \label{fig:sketch_thermobar}
  \end{subfigure}\hfill
  \begin{subfigure}[t]{0.5\textwidth}
    \centering
    \includegraphics[scale=0.13,trim={4cm 4cm 6cm 3cm},clip]{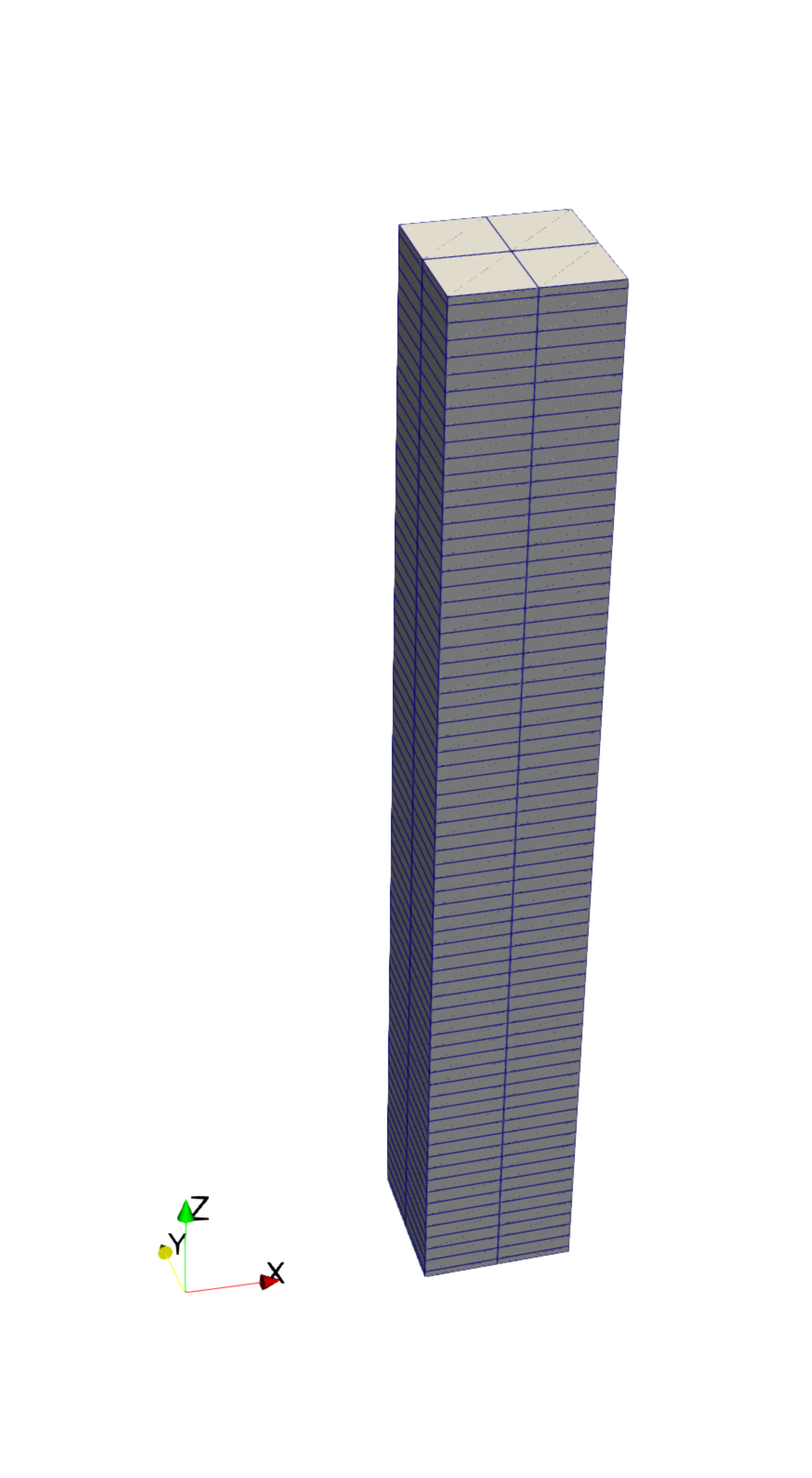}
    \caption{}
    \label{fig:domain_thermobar}
  \end{subfigure}
  \caption{Thermo-poroelasticity verification problem in experiment \ref{ssec:thermo_bar}. (a) Schematic of the 1D saturated porous column subjected to a compressive stress load ($\textbf{F}$) applied on the top boundary. The bottom boundary is fixed (Dirichlet condition imposed on the momentum balance Eq.~\eqref{eq:balance_momentum}), and zero normal displacement (roller conditions) is imposed on the lateral boundaries. (b) Corresponding three-dimensional domain discretized with hexahedral elements.}
  \label{fig:thermobar_setup}
\end{figure}

\begin{figure}
  \centering
  \begin{subfigure}[t]{\textwidth}
    \centering
    \includegraphics[scale=0.58,trim={0cm 0cm 0cm 0cm},clip]{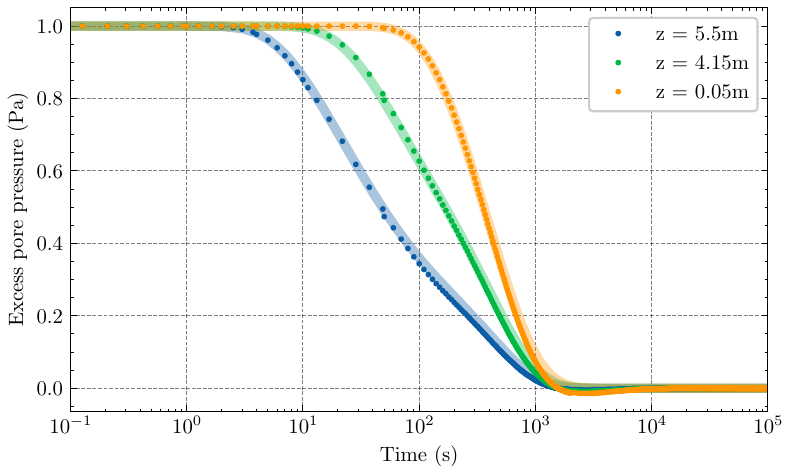}
    \caption{}
    \label{fig:thermobar_pres}
  \end{subfigure}\hfill
  \begin{subfigure}[t]{\textwidth}
    \centering
    \includegraphics[scale=0.58,trim={0cm 0cm 0cm 0cm},clip]{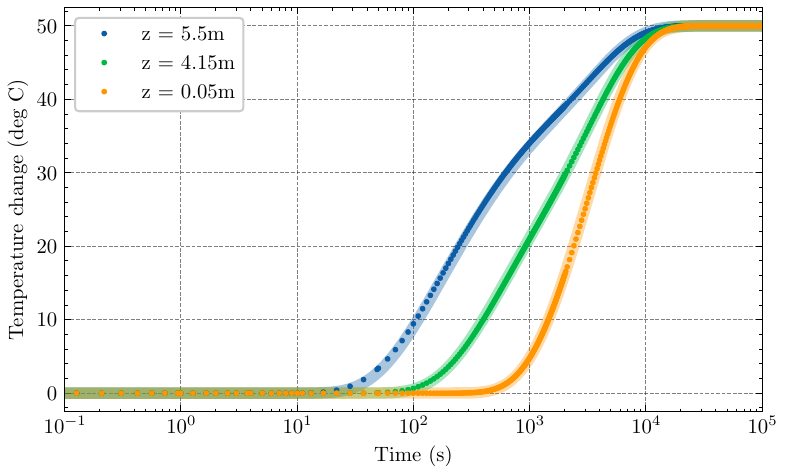}
    \caption{}
    \label{fig:thermobar_temp}
  \end{subfigure}\hfill
  \begin{subfigure}[t]{\textwidth}
    \centering
    \includegraphics[scale=0.58,trim={0cm 0cm 0cm 0cm},clip]{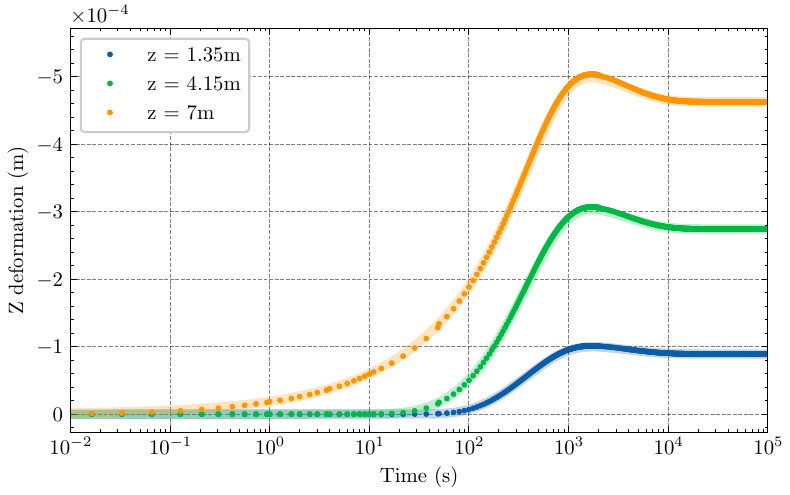}
    \caption{}
    \label{fig:thermobar_disp}
  \end{subfigure}
  \caption{Temporal evolution of (a) pore pressure, (b) temperature, and (c) vertical deformation at three locations in the one-dimensional thermo-poroelastic problem~(\ref{ssec:thermo_bar}). Symbols denote the PFLOTRAN solution and the solid lines denote the semi-analytical solution of \cite{Bai_2005}.}
  \label{fig:thermobar_comparison}
\end{figure}

The thermo-poroelastic formulation is verified using the one-dimensional problem described by \cite{Bai_2005} and \cite{gao_2019}. A homogeneous, fully saturated porous column is initialized with a uniform state (initial conditions for the mass balance Eq.~\eqref{eq:mass_balance} and the energy balance Eq.~\eqref{eq:energy_balance} are $P_i = 101325$ Pa and $T_i = 0$\degree C, respectively). On the top boundary, a compressive stress load of 1 Pa is suddenly applied and the temperature is raised to $T_{BC,top}=50$\degree C, which subsequently diffuses into the domain. All other boundaries are set impermeable (no-flux) and thermally insulating (zero heat flux) conditions such that no-flux conditions are prescribed to the mass balance Eq.~\eqref{eq:mass_balance} and the energy balance Eq.~\eqref{eq:energy_balance}; the mechanical load is a traction boundary condition for the momentum balance Eq.~\eqref{eq:balance_momentum}. The loading produces an initial undrained pore-pressure increase, followed by pressure diffusion as the fluid drains from top boundary, while the temperature rises gradually due to thermal diffusion. A sketch of the conceptual model and the element mesh made up of hexahedrons are shown in Fig. \ref{fig:sketch_thermobar} and \ref{fig:domain_thermobar}, respectively. Table \ref{tab:thermoelastic_params} provides the parameters used in this verification. Fig. \ref{fig:thermobar_comparison} shows the pore pressure, temperature, and displacement profiles at three depths with the analytical solution of \cite{Bai_2005}.

\begin{table}[t]
\centering
\caption{Setup and input parameters for thermo-poroelastic consolidation in experiment \ref{ssec:thermo_bar}.}
\label{tab:thermoelastic_params}
\begin{tabularx}{\linewidth}{X r l}
%\end{tabularx}
\toprule
Parameter & Value & Unit \\
\midrule
Domain length, width, height ($L_x$, $L_y$, $L_z$) & 0.1, 0.1, 7 & m \\
Porosity ($\phi$) & 0.2 & -- \\
Young's modulus ($E$) & 6000 & Pa \\
Poisson's ratio ($\nu$) & 0.40 & -- \\
Solid volumetric heat capacity ($\rho C$) & \num{167.2e3} & J/(m$^{3}\,$\degree C) \\
Thermal conductivity ($k_T$) & 836 & J/(m\,s\,\degree C) \\
Thermal expansion coefficient ($\alpha$) & \num{1e-7} & (1./\degree C) \\
Permeability ($k$) & 4.e-9 & $m^2$ \\
Viscosity ($\mu$) & 0.001 & Pa s \\
Biot's coefficient ($b$) & 1 & -- \\
Initial temperature ($T_{i}$) & 0 & \degree C \\
Initial Pressure ($P_{i}$) & 101325 & Pa \\
Top surface temperature ($T_{top, BC}$) & 50 &  \degree C \\
Applied compressive stress load ($\mathbf{F}$) & 1 & Pa \\
Number of elements & 284 & -- \\
Number of Voronoi cells & 70 & -- \\
\bottomrule
\end{tabularx}
\end{table}

\subsection{Pressurized-crack} \label{ssec:crack}
\begin{figure}[htbp]
  \begin{subfigure}[t]{0.4\textwidth}
    \centering
    \includegraphics[scale=0.5,trim={0cm 0cm 1cm 0cm},clip]{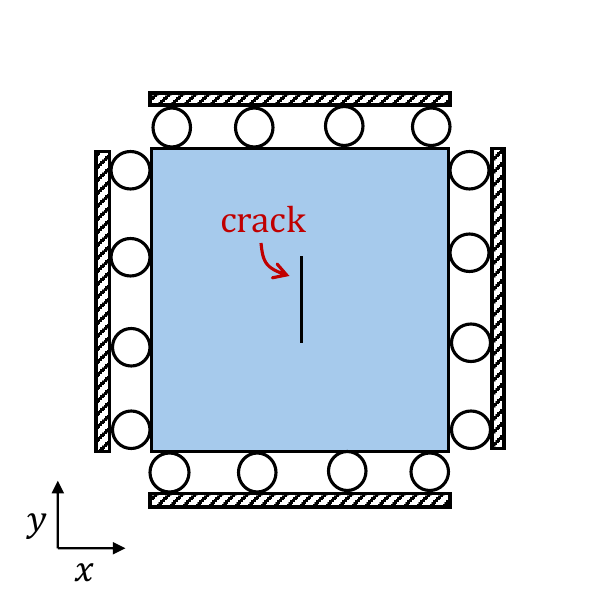}
    \caption{}
    \label{fig:sketch_crack}
  \end{subfigure}\hfill
  \begin{subfigure}[t]{0.6\textwidth}
      \centering
    \includegraphics[scale=0.18,trim={0cm 0cm 0.5cm 3cm},clip]{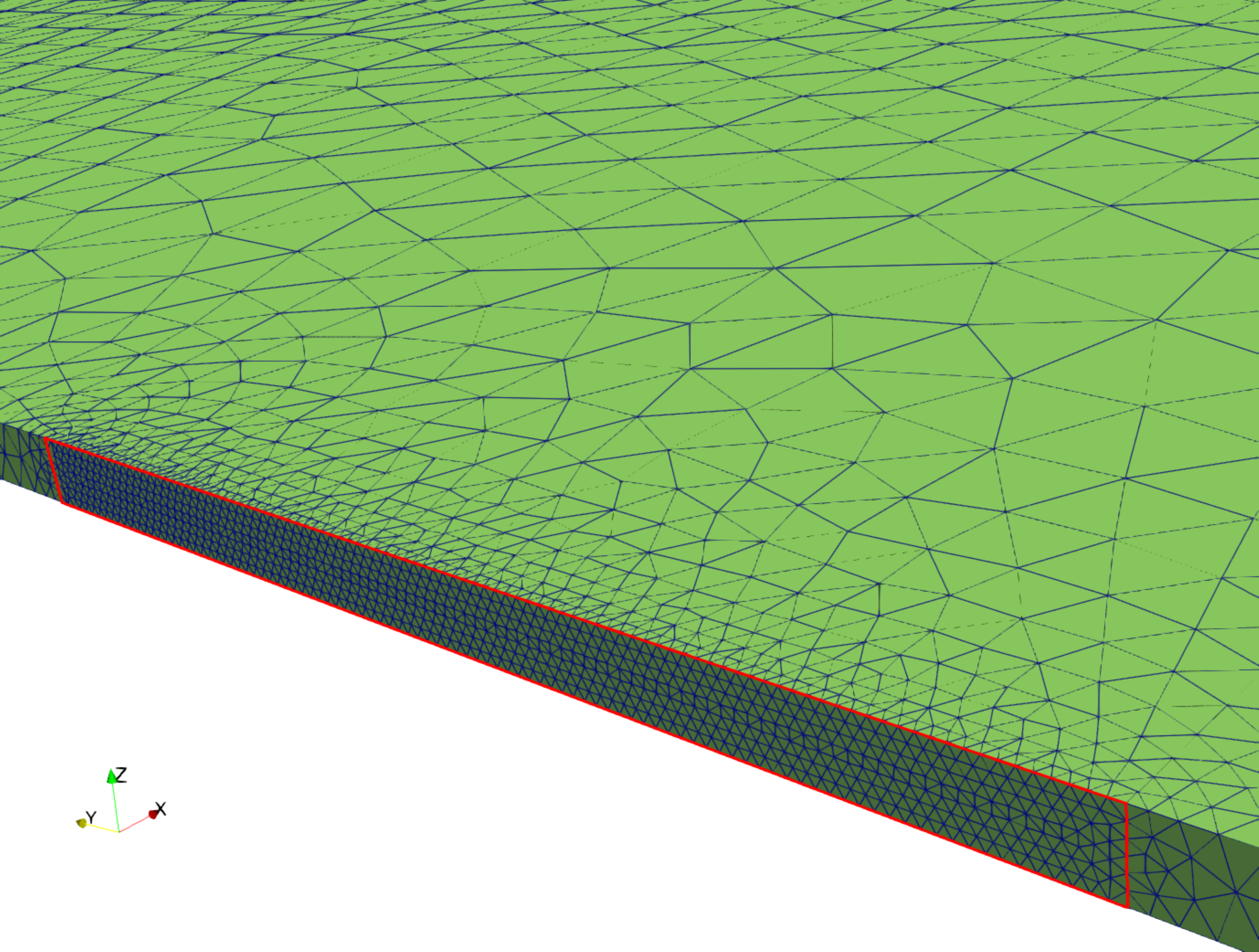}
    \caption{}
    \label{fig:mesh_crack}
  \end{subfigure}
  \begin{subfigure}[t]{\textwidth}
    \centering
    \includegraphics[scale=0.18,trim={0cm 0cm 0cm 8cm},clip]{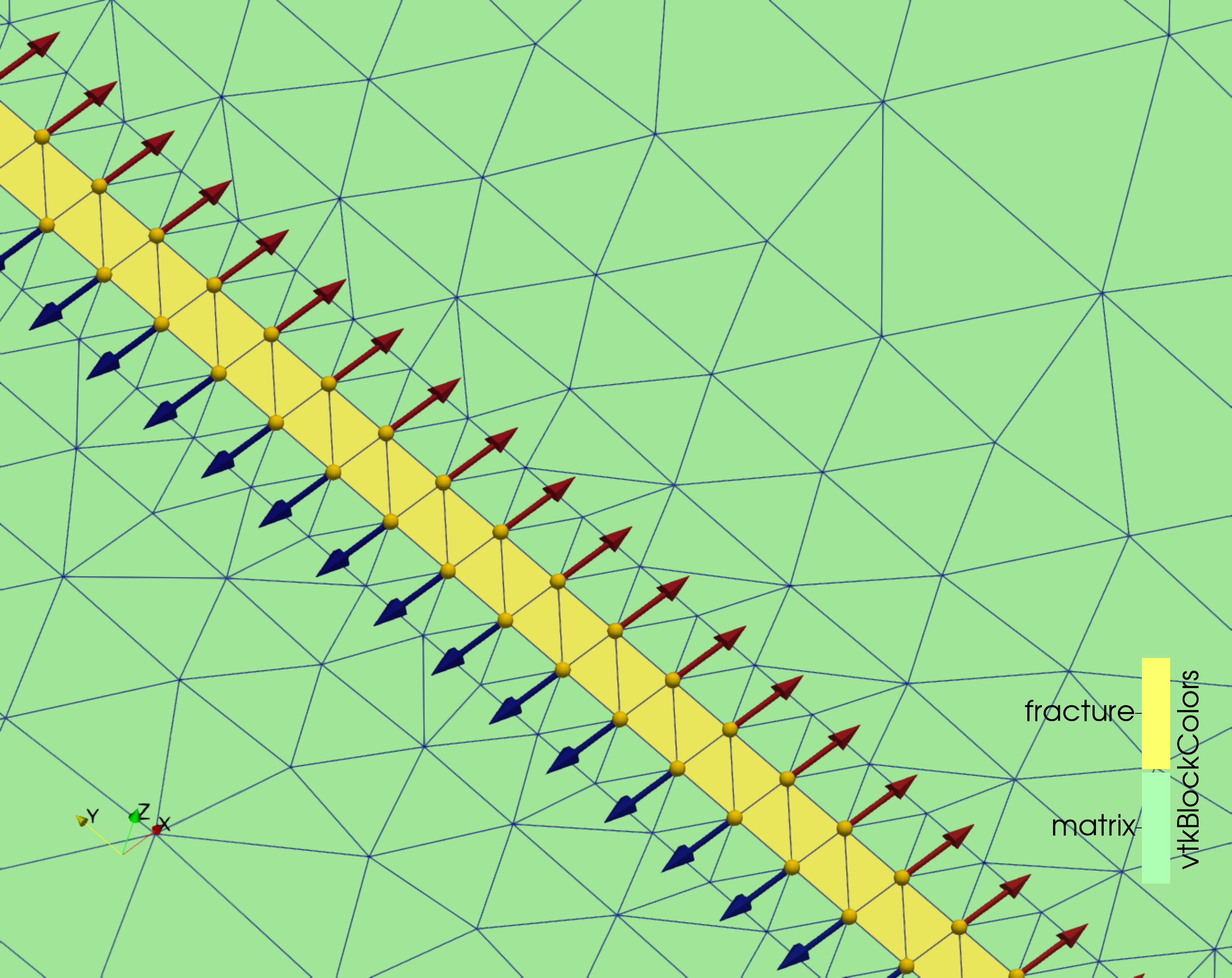}
    \caption{}
    \label{fig:frac_view}
  \end{subfigure}
  \caption{Pressurized-crack verification problem (Experiment \ref{ssec:crack}). (a) 2D schematic of the problem setup, showing a central crack embedded in the domain with zero normal displacement prescribed to the outer boundaries (Dirichlet condition imposed on the momentum balance Eq.~ \eqref{eq:balance_momentum}); plane-strain conditions are assumed in the out-of-plane (z) direction. (b) Purely mechanical model, which exploits symmetry to represent only half of the domain. The close-up view shows the 3D mesh near the crack interface (red outline), where the mechanical boundary conditions are applied. (c) Hydro-mechanical model, in which the fracture is explicitly assigned as a region (yellow) embedded within the surrounding matrix (green). The close-up top view shows the element mesh near the fracture, with fracture nodes highlighted (yellow circles). To preserve the pressure gradient with flow–mechanics node mapping in PFLOTRAN, the fracture must be at least one element wide to ensure accurate nodal displacements (the fracture pressure is imposed on the mass balance Eq.~\eqref{eq:mass_balance}). Arrows indicate the direction of displacement experienced by each fracture node.}
  \label{fig:setup_crack}
\end{figure}

In the classical two-dimensional pressurized-crack problem \citep{Sneddon1946, Sneddon_1946_b}, we consider a planar crack in an infinite, homogeneous, linear-elastic medium subjected to a uniform internal pressure applied on the crack surface. Fig. \ref{fig:sketch_crack} shows the sketch of the problem. The parameters that describe the problem are given in Table \ref{tab:crack_params}. The crack opening (displacement field) is commonly used as a benchmark for verifying numerical implementations for hydraulic fracturing and pressurized fracture mechanics in linear elastic models \citep{Sneddon_1946_b, adachi_2007, Anderson_2017}. Here, we employ two approaches to verify the formulation for mechanical loading and hydraulic loading. In the purely mechanical setting, the displacement solution is recovered by applying a mechanical load where the solution is directly compared to the analytical displacement profile, Fig. \ref{fig:mesh_crack} (the mechanical load is applied as a traction boundary condition in the momentum balance Eq.~\eqref{eq:balance_momentum}). In the hydro-mechanical setting, the displacement solution is achieved by imposing an equivalent pressure gradient between the fracture region and the surrounding rock, so that the mechanical solver via Biot's coupling term reproduces the equivalent loading condition (pressure condition applied in the mass balance Eq.~\eqref{eq:mass_balance}).

\begin{figure}
\centering
\includegraphics[scale=0.65,trim={0cm 0cm 0cm 0cm},clip]{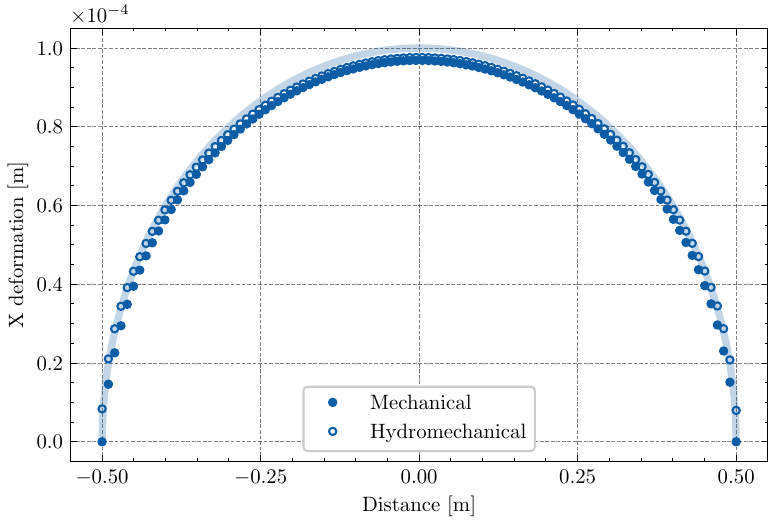}
\caption{Crack-opening displacement profile along the crack interface for the pressurized-crack problem in experiment~\ref{ssec:crack}. The solid line denotes the analytical Sneddon solution, and the symbols denote the PFLOTRAN results for the purely mechanical and hydro-mechanical setups.}\label{fig:Sneddon_comparison}
\end{figure}

In the hydro-mechanical problem, we ensure that the pressure gradient is captured by employing two control volumes (flow mesh), which are mapped to one element of the fracture region in the mechanics domain, Fig. \ref{fig:frac_view}. This approach ensures consistency is preserved across the coupled simulation and that the effective loading conditions in the hydro-mechanical model are applied properly (pressure conditions for the mass balance Eq.~\eqref{eq:mass_balance} mapped to the momentum balance Eq.~\eqref{eq:balance_momentum}). Numerical results from both the purely mechanical and the hydro-mechanical setups closely follow the analytical Sneddon solution (Fig. \ref{fig:Sneddon_comparison}), reproducing the symmetric opening profile with maximum displacement at the crack center and closure toward the tips. The relative $L_2$-norm errors for the mechanical and hydro-mechanical approaches are 0.041 and 0.026, respectively (Table \ref{tab:l2_err}).
\begin{table}[t]
\centering
\caption{Setup and input parameters for pressurized-crack problem in experiment \ref{ssec:crack}.}
\label{tab:crack_params}
\begin{tabularx}{\linewidth}{X r l}
\toprule
Parameter & Value & Unit \\
\midrule
Domain length, width, height ($L_x$, $L_y$, $L_z$) & 20, 20, 0.06 & m \\
Crack length ($L_f$) & 1 & m \\
Young's modulus $(E)$ & 10 & GPa \\
Poisson's ratio $(\nu)$ & 0 & -- \\
Applied load ($p_w$, $\Delta p$) & 1 & MPa \\
Element edge length on crack interface $(h)$ & 1 & cm \\
Fracture aperture (hydromechanics) & 1 & cm \\
Number of elements (mechanics, hydromechanics) & 53k, 58k & -- \\
Number of Voronoi cells (mechanics, hydromechanics) & ~17k, 16k & -- \\
\bottomrule
\end{tabularx}
\end{table}

\subsection{Circular borehole} \label{ssec:borehole}
Stress concentrations around drilled boreholes are critical for understanding and evaluating subsurface formations in various field experiments. Here we verify the robustness of the numerical implementation of the mechanics solver to capture stress distributions around a circular borehole for future work involving simulation of coupled subsurface processes.

\begin{figure}
  \centering
  \begin{subfigure}{0.52\textwidth}
    \centering
    \includegraphics[scale=0.5]{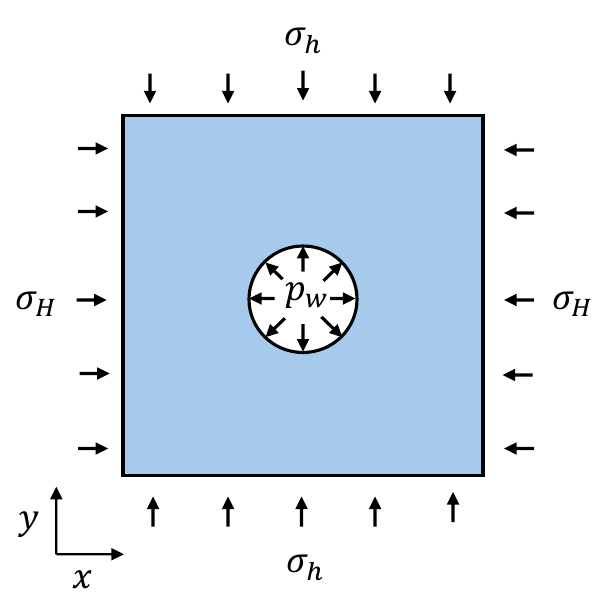}
    \caption{}
    \label{fig:kirsch_schematic}
  \end{subfigure}\hfill
  \begin{subfigure}{0.48\textwidth}
    \centering
    \includegraphics[scale=0.2,trim={5.5cm 4cm 17cm 25cm},clip]{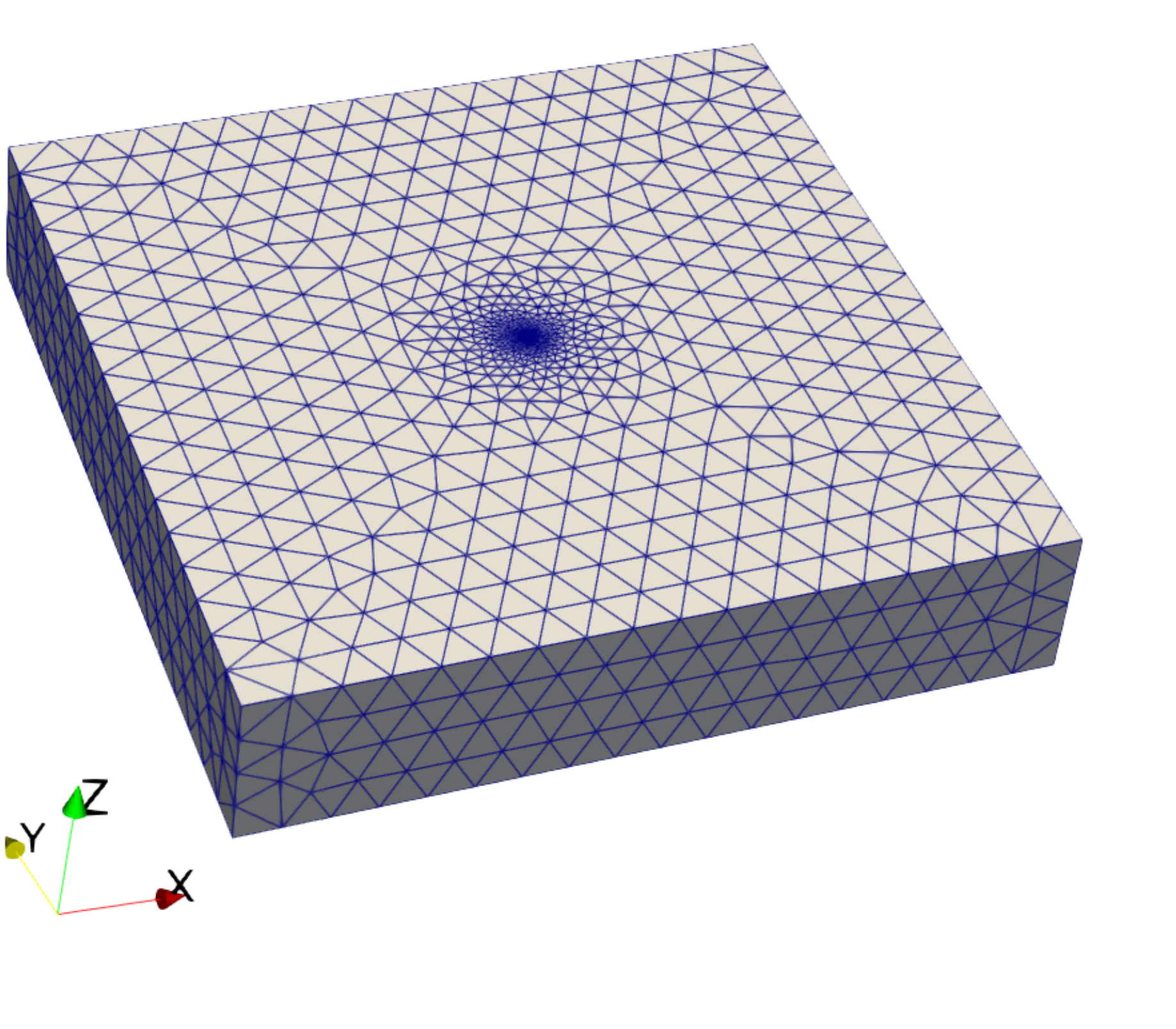}
    \caption{}
    \label{fig:kirsch_mesh}
  \end{subfigure}
  \caption{(a) Sketch of Kirsch's setup in experiment \ref{ssec:borehole} for the circular borehole domain subjected to biaxial far-field stresses ($\sigma_H$ and $\sigma_h$) and applied internal cavity pressure ($p_w$), applied as traction boundary in the momentum balance Eq.~ \eqref{eq:balance_momentum}). (b) Corresponding three-dimensional element mesh used for the plane-strain model. Roller boundary conditions are applied on the top and bottom (z-normal) faces (zero normal displacement in Eq.~\eqref{eq:balance_momentum}). The mesh is locally refined near the borehole to capture the stress concentration.}
  \label{fig:kirsch_setup}
\end{figure}

\begin{figure}
\centering
  \begin{subfigure}{\textwidth}
  \centering
    \includegraphics[scale=0.58,trim={0cm 0.15cm 0cm 0.05cm},clip]{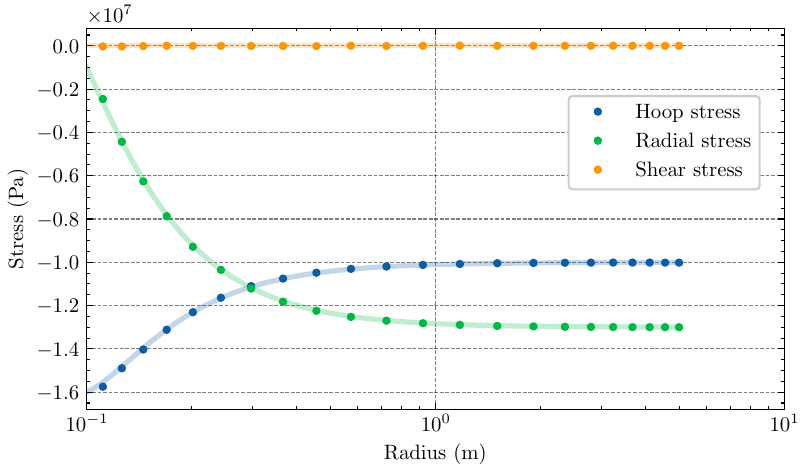}
    \caption{}
    \label{fig:kirsch_theta0}
  \end{subfigure}\hfill
  \begin{subfigure}{\textwidth}
  \centering
    \includegraphics[scale=0.58,trim={0cm 0.15cm 0cm 0.05cm},clip]{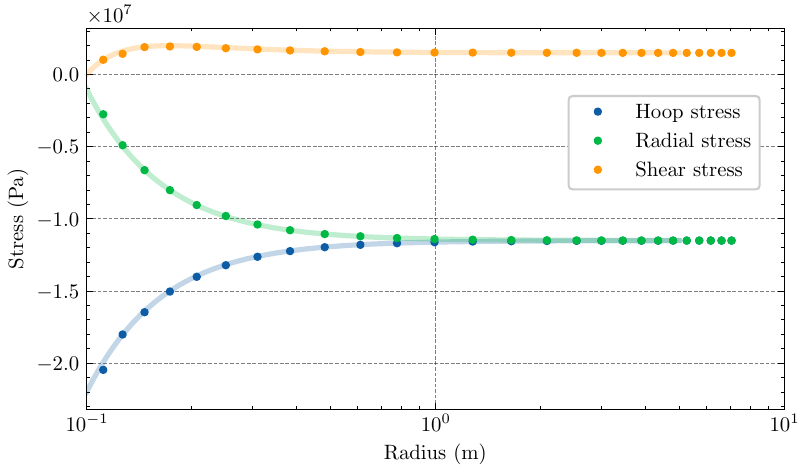}
    \caption{}
    \label{fig:kirsch_theta45}
  \end{subfigure}\hfill
  \begin{subfigure}{\textwidth}
  \centering
    \includegraphics[scale=0.58,trim={0cm 0.15cm 0cm 0.05cm},clip]{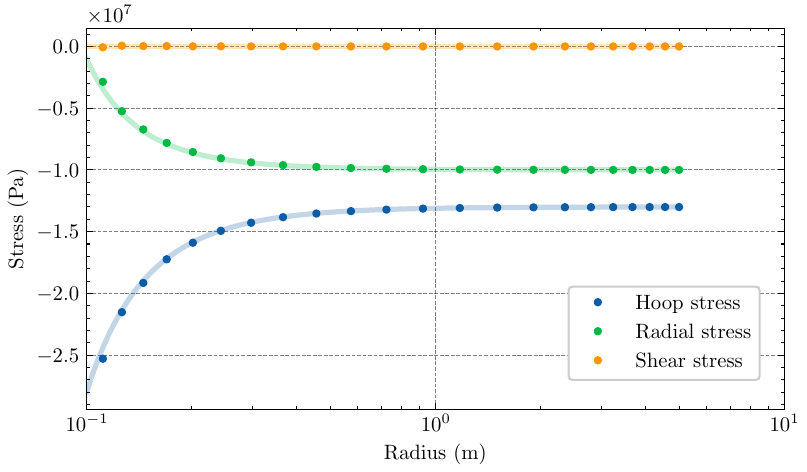}
    \caption{}
    \label{fig:kirsch_theta90}
  \end{subfigure}
\caption{Kirsch benchmark results for experiment~\ref{ssec:borehole}. Symbols denote the PFLOTRAN solution and the solid lines denote the analytical solution. The radial, hoop, and shear stress components are plotted as functions of radial distance from the cavity wall for (a) $\theta = 0^\circ$, (b) $\theta = 45^\circ$, and (c) $\theta = 90^\circ$.}\label{fig:kirsch_stresses}
\end{figure}

The numerical results are verified against Kirsch's analytical solution \citep{kirsch_1898, Jaeger_2007, Zoback_2007} (Fig. \ref{fig:kirsch_schematic}). The benchmark provides a solution for the stress distribution around a circular borehole subjected to biaxial far-field stresses, $\sigma_H$ and $\sigma_h$, and internal borehole pressure ($p_w$). Both far-field stresses and internal pressure act as boundary conditions on the momentum balance Eq.~\eqref{eq:balance_momentum}. The three-dimensional domain is shown in Fig. \ref{fig:kirsch_mesh} and the parameters employed in the experiment are described in Table \ref{tab:kirsch_params}. Fig. \ref{fig:kirsch_stresses} shows the numerical results and Kirsch's solution.

The stress components are plotted as functions of radial distance from the borehole wall for three angles, $\theta = 0 \degree$, $45 \degree$, and $90 \degree$. The first angle aligns with the maximum far-field stress and yields the largest radial compressive stresses; the second exhibits nonzero shear stresses due to far-field stress anisotropy; and the last aligns with the minimum far-field stress. The relative $L_2$-norm errors for shear stress, hoop stress, and radial stress are below 0.03, 0.02, and 0.02, respectively (Table~\ref{tab:l2_err}).
\begin{table}[t]
\centering
\caption{Setup and input parameters for circular borehole (Kirsch's) problem in experiment \ref{ssec:borehole}.}
\label{tab:kirsch_params}
\begin{tabularx}{\linewidth}{X r l}
\toprule
Parameter & Value & Unit \\
\midrule
Domain length, width, height ($L_x$, $L_y$, $L_z$) & 10, 10, 2 & m \\
Borehole radius ($r_w$) & 10 & cm \\
Young's modulus $(E)$ & 1 & GPa \\
Poisson's ratio $(\nu)$ & 0.1 & -- \\
Far-field maximum stress $(\sigma_H)$ & -13 & MPa \\
Far-field minimum stress $(\sigma_h)$ & -10 & MPa \\
Internal applied pressure $(p_w)$ & 1 & MPa \\
Measured angles from borehole axis $(\theta)$ & 0, 45, 90 & degrees \\
Number of elements & 287k & -- \\
Number of Voronoi cells & 56k & -- \\
\bottomrule
\end{tabularx}
\end{table}

\section{Discussion}\label{sec:discussion}
This manuscript is intended primarily as a verification study for the PFLOTRAN THM implementation. The benchmark suite demonstrates that the sequential fixed-stress formulation reproduces analytical or semi-analytical responses for one-dimensional consolidation, two-dimensional poroelastic stress redistribution, thermo-poroelastic response, discontinuity loading, and borehole stress concentration. In that sense, the results establish that the implemented capability recovers accepted reference behavior for the classes of problems tested here.

The discontinuity and borehole examples play a complementary role in the verification suite. The pressurized-crack problem verifies that the mapping between flow and geomechanics degrees of freedom reproduces the expected Sneddon crack-opening response for the tested geometry and resolution. The Kirsch benchmark verifies the geomechanics stress response around a borehole for several angular directions. Together, these cases broaden the verification coverage beyond one-dimensional consolidation and show that the implementation behaves consistently for mechanically important stress and displacement observables.

\section{Conclusion}\label{sec:conclusion}
The verification results presented here show that the PFLOTRAN THM capability reproduces analytical or semi-analytical benchmark responses for pressure diffusion, thermo-poroelastic evolution, deformation, discontinuity loading, and borehole stress redistribution within the scope of the problems considered. The implementation leverages the well-established finite-volume framework within PFLOTRAN for mass and energy balance and extends it with a finite-element approach to solve the quasi-static momentum balance equation. The coupling is achieved by using a non-iterative fixed-stress split strategy. Furthermore, we demonstrate the proposed treatment for describing discontinuities (i.e. presence of fractures or cracks) based on mapping between mechanical and flow degrees of freedom suitable for PFLOTRAN implementation. The benchmark suite therefore provides a verification-oriented foundation for future THM studies in geologic porous media and engineered energy applications.

\backmatter

%\bmhead{Supplementary information}
%\PLACEHOLDER{Data repository link or archive DOI}. \PLACEHOLDER{Repository accession or archive identifier}. Code is available at \PLACEHOLDER{GitHub or Zenodo URL with tagged version/release}.

\bmhead{Acknowledgements}
This research was supported by CUSSP (Center for Understanding Subsurface Signals and Permeability), funded by the U.S. Department of Energy (DOE), Office of Science under FWP 81834. PNNL is operated for the DOE by Battelle Memorial Institute under contract DE-AC05-76RL01830. This paper describes objective technical results and analysis. Any subjective views or opinions that might be expressed in the paper do not necessarily represent the views of the U.S. Department of Energy or the United States Government.

\begin{comment}
\begin{itemize}
\item Funding
\item Conflict of interest/Competing interests (check journal-specific guidelines for which heading to use)
\item Ethics approval and consent to participate
\item Consent for publication
\item Data availability
\item Materials availability
\item Code availability
\item Author contribution
\end{itemize}

\noindent
If any of the sections are not relevant to your manuscript, please include the heading and write `Not applicable' for that section.
\end{comment}
%%===================================================%%
%% For presentation purpose, we have included        %%
%% \bigskip command. Please ignore this.             %%
%%===================================================%%
\begin{comment}
\begin{flushleft}%
Editorial Policies for:

\bigskip\noindent
Springer journals and proceedings: \url{https://www.springer.com/gp/editorial-policies}

\bigskip\noindent
Nature Portfolio journals: \url{https://www.nature.com/nature-research/editorial-policies}

\bigskip\noindent
\textit{Scientific Reports}: \url{https://www.nature.com/srep/journal-policies/editorial-policies}

\bigskip\noindent
BMC journals: \url{https://www.biomedcentral.com/getpublished/editorial-policies}
\end{flushleft}
\end{comment}

\begin{appendices}
\section{Tabulated relative L2 errors}\label{appen:l2norm}

% Optional: state the definition you use
\noindent The relative $L_2$ error is defined as
\begin{equation}
\varepsilon_{L_2} = \frac{\|x_h - x\|_{L_2(\Omega)}}{\|x\|_{L_2(\Omega)}},
\label{eq:l2_err}
\end{equation}
%\[
%\varepsilon_{L_2} = \frac{\|x_h - x\|_{L_2(\Omega)}}{\|x\|_{L_2(\Omega)}} .
%\]
where $x_h$ and $x$ denote the numerical solution and the corresponding analytical solution, respectively, for the variable of interest (e.g., pressure, temperature, displacement, strain, or stress).

\begin{table}[tbp]
\centering
\caption{Relative $L_2$ errors for the numerical experiments.}
\label{tab:l2_err}
\scriptsize
\begin{tabular}{llll}
\toprule
\textbf{Experiment} & \textbf{Evaluation} & {\textbf{Rel. $L_2$ error}} & \textbf{Variable} \\
\midrule
\multirow{2}{*}{Terzaghi's problem (\ref{ssec:1d_consolidation})}
  & $t=0.005\,\mathrm{s}$ & \num{1.2e-3} & pressure \\
  & $t=0.05\,\mathrm{s}$ & \num{3.7e-4} & pressure \\
  & $t=0.2\,\mathrm{s}$ & \num{4.7e-4} & pressure \\
  & $t=0.5\,\mathrm{s}$ & \num{1.4e-3} & pressure \\
  & $t=0.005\,\mathrm{s}$ & \num{9.0e-3} & displacement \\
  & $t=0.05\,\mathrm{s}$ & \num{7.3e-4} & displacement \\
  & $t=0.2\,\mathrm{s}$ & \num{5.8e-4} & displacement \\
  & $t=0.5\,\mathrm{s}$ & \num{4.7e-4} & displacement \\
\midrule
\multirow{2}{*}{Schiffman's problem (\ref{ssec:1d_consolidation})}
  & $t=0.001\,\mathrm{s}$ & \num{3.4e-2} & pressure \\
  & $t=0.002\,\mathrm{s}$ & \num{1.4e-2} & pressure \\
  & $t=0.003\,\mathrm{s}$ & \num{9.4e-3} & pressure \\
  & $t=0.05\,\mathrm{s}$  & \num{2.7e-3} & pressure \\
  & $t=0.2\,\mathrm{s}$   & \num{1.6e-3} & pressure \\
  & $t=0.001\,\mathrm{s}$ & \num{3.2e-1} & strain \\
  & $t=0.002\,\mathrm{s}$ & \num{1.7e-1} & strain \\
  & $t=0.003\,\mathrm{s}$ & \num{1.3e-1} & strain \\
  & $t=0.05\,\mathrm{s}$  & \num{2.3e-2} & strain \\
  & $t=0.2\,\mathrm{s}$   & \num{9.5e-3} & strain \\
\midrule
\multirow{3}{*}{Mandel's problem (\ref{ssec:mandel})}
  & $t=0.01\,\mathrm{s}$ & \num{1.0e-2} & pressure \\
  & $t=0.1\,\mathrm{s}$ & \num{2.3e-2} & pressure \\
  & $t=0.5\,\mathrm{s}$ & \num{2.3e-2} & pressure \\
  & $t=5\,\mathrm{s}$ & \num{1.9e-2} & pressure \\
  & $t=0.01\,\mathrm{s}$ & \num{2.5e-2} & horizontal displacement \\
  & $t=0.1\,\mathrm{s}$ & \num{6.1e-2} & horizontal displacement \\
  & $t=0.5\,\mathrm{s}$ & \num{9.5e-2} & horizontal displacement \\
  & $t=5\,\mathrm{s}$ & \num{9.0e-2} & horizontal displacement \\
  & $t=0.01\,\mathrm{s}$ & \num{1.9e-2} & vertical displacement \\
  & $t=0.1\,\mathrm{s}$ & \num{5.5e-2} & vertical displacement \\
  & $t=0.5\,\mathrm{s}$ & \num{9.5e-2} & vertical displacement \\
  & $t=5\,\mathrm{s}$ & \num{4.7e-2} & vertical displacement \\
\midrule
\multirow{3}{*}{Thermo-poroelastic bar (\ref{ssec:thermo_bar})}
  & $z=5.5\,\mathrm{m}$ & \num{2.7e-1} & pressure \\
  & $z=4.15\,\mathrm{m}$ & \num{2.4e-1} & pressure \\
  & $z=0.05\,\mathrm{m}$ & \num{2.1e-1} & pressure \\
  & $z=5.5\,\mathrm{m}$ & \num{4.1e-3} & temperature \\
  & $z=4.15\,\mathrm{m}$ & \num{8.3e-4} & temperature \\
  & $z=0.05\,\mathrm{m}$ & \num{1.0e-3} & temperature \\
  & $z=1.35\,\mathrm{m}$ & \num{3.6e-2} & displacement \\
  & $z=4.15\,\mathrm{m}$ & \num{3.6e-2} & displacement \\
  & $z=7\,\mathrm{m}$ & \num{3.6e-2} & displacement \\
\midrule
\multirow{2}{*}{Pressurized crack (\ref{ssec:crack})}
  & mechanical & \num{4.1e-2} & displacement \\
  & hydro-mechanical & \num{2.6e-2} & displacement \\
\midrule
\multirow{3}{*}{Circular borehole (\ref{ssec:borehole})}
  & $\theta=0^\circ$ & \num{4.6e-3} & hoop stress \\
  & $\theta=45^\circ$ & \num{8.3e-3} & hoop stress \\
  & $\theta=90^\circ$ & \num{1.3e-2} & hoop stress \\
  & $\theta=0^\circ$ & \num{3.8e-3} & radial stress \\
  & $\theta=45^\circ$ & \num{6.3e-3} & radial stress \\
  & $\theta=90^\circ$ & \num{1.5e-2} & radial stress \\
  & $\theta=0^\circ$ & -$^{\dagger}$ & shear stress \\
  & $\theta=45^\circ$ & \num{2.6e-2} & shear stress \\
  & $\theta=90^\circ$ & -$^{\dagger}$ & shear stress \\
\bottomrule
\end{tabular}
\par\smallskip
\footnotesize
$^{\dagger}$ The analytical shear stress is identically zero at $\theta=0^\circ$ and $\theta=90^\circ$ by symmetry; a relative error is therefore not reported.\\
\emph{Note:} For experiment~\ref{ssec:thermo_bar}, the reported errors are time-history metrics evaluated at the stated sensor locations; this table therefore combines profile-based and time-history-based verification quantities in order to keep the benchmark evidence self-contained in the manuscript.
\end{table}

\end{appendices}

%%===========================================================================================%%
%% If you are submitting to one of the Nature Portfolio journals, using the eJP submission   %%
%% system, please include the references within the manuscript file itself. You may do this  %%
%% by copying the reference list from your .bbl file, paste it into the main manuscript .tex %%
%% file, and delete the associated \verb+\bibliography+ commands.                            %%
%%===========================================================================================%%

\bibliography{bibliography}% common bib file
%% if required, the content of .bbl file can be included here once bbl is generated
%%\input sn-article.bbl

\end{document}